\documentclass[12pt]{iopart}

\usepackage{iopams}
\usepackage{graphicx}
\usepackage{psfrag}
\usepackage{epstopdf}
\usepackage{color}
\usepackage{hyperref}

\providecommand{\prob}{{\rm P}}
\providecommand{\openone}{\mbox{{\small 1}$\!\!$1}}
\newcommand{\qC}[3]{\left[\!\begin{array}{c}#1\\#2\end{array}\!\right]_{#3}}

\begin{document}

\title{Matrix product solution of the multispecies partially asymmetric exclusion process}

\author{S. Prolhac$^{1}$, M. R. Evans$^{2}$, K. Mallick$^{1}$}

\address{$^{1}$Institut de Physique Th\'eorique\\CEA, IPhT, F-91191
 Gif-sur-Yvette, France\\CNRS, URA 2306, F-91191 Gif-sur-Yvette,
 France\\ $^2$ SUPA, School of Physics and Astronomy,\\University of
 Edinburgh, Mayfield Road, Edinburgh EH9 3JZ, UK}

\ead{sylvain.prolhac@cea.fr,m.evans@ed.ac.uk,kirone.mallick@cea.fr}

\paragraph{\bf Abstract}
We find the exact solution for the stationary state measure of the
partially asymmetric exclusion process on a ring with multiple species
of particles.  The solution is in the form of a matrix product
representation where the matrices for a system of $N$ species are
defined recursively in terms of the matrices for a system of $N{-}1$
species.  A complete proof is given, based on the quadratic relations
verified by these matrices. This matrix product construction is
interpreted in terms of the action of a transfer matrix.
\medskip



\section{Introduction}
Particles hopping on a one-dimensional lattice with hard-core
exclusion interactions provide a simple framework for the study of
interacting many-body systems \cite{Liggett85,Spohn91}. In particular,
when the particle hopping is asymmetric, a macroscopic particle
current results and the system attains a nonequilibrium steady state
(NESS) in which detailed balance is not satisfied (see \cite{BE07} for
a recent review).

A fundamental example of such a system is the Asymmetric Simple
Exclusion Process (ASEP) \cite{D98,GM06}. Here particles attempt hops
to the right neighbour site with unit rate and to the left neighbour
site with rate $q$. The hop is carried out when the destination site
is empty. The special case $q=0$ is known as the Totally Asymmetric
Simple Exclusion Process (TASEP). With periodic boundary conditions
the NESS of the ASEP has a very simple form (all allowed
configurations of particles are equally likely) yet dynamical
properties such as diffusion of a tagged particle
\cite{DEM93,DM97,PM08,P08}, and large deviations of the current
\cite{DL98} have proved to be non-trivial. When open boundary
conditions, where particles attempt to enter and exit at the left and
the right boundary, are used instead of periodic, the NESS takes on a
non-trivial form. It may be represented by a matrix product state
\cite{DEHP93} in which the steady state probabilities for each
configuration are obtained from products of two matrices $D$ and $E$
according to whether each site is occupied or empty in the
configuration. These matrices obey a quadratic algebra which provides
a motif from which all the steady state probabilities may be
generated. This quadratic algebra is related to the q-deformed
harmonic oscillator \cite{DM97,ER96,Sasamoto99,BECE00,EB02,MS97}.

A generalization of the ASEP is to the case of two species of
particles. A well studied model is that of usual (first-class) and
second class particles. In this case both first and second-class
particles hop to the right with unit rate and to the left with rate
$q$. However if the site to the right of a first-class particle is
occupied by a second-class particle the first and second-class
particle exchange places with rate $1$. Conversely, if the site to the
left of a first-class particle is occupied by a second-class particle
they can exchange with rate $q$. Thus a second-class particle behaves
as a hole from the point of view of a first-class particle but behaves
as a particle from the point of view of a hole.  When $q$ is lower
than $1$ the second class particles will move forwards in an
environment of a low density of first class particles but backwards in
a high density environment, therefore the introduction of a second
class particle is a useful tool in the study of the microscopic
structure of shocks \cite{BCFN89,FKS,Pablo91,DJLS93,M96}. The
stationary state of a periodic system containing second and
first-class particles has been obtained using the matrix product
formulation by Derrida {\it et al.}  \cite{DJLS93}. Here the three
matrices $D$, $E$ and $A$ (corresponding to first-class particles,
second-class particles and holes respectively) obey a quadratic
algebra closely related to that of \cite{DEHP93}. Quadratic algebras
for two species exclusion process have been further explored
\cite{Karimipour99,KK00,AHR98} and a general classification of
quadratic algebras has been made in \cite{IPR01}. The range of matrix
product states and quadratic algebras relevant to exclusion process
and other stochastic systems have been summarized in \cite{BE07}.

A natural generalization of the two species case of first and second
class particles is to the multispecies process where there is a
hierarchy amongst the different species. That is, for a system with
$N$ species, the $N$th class particles are treated by all other
classes of particle as holes, $(N{-}1)$th class particles treat $N$th
class particles as holes but are themselves treated as holes by first
class, second class,.... $(N{-}2)$th class particles, and so on, up to
the first class particles which treat all other species as holes. We
refer to this model as the $N$-ASEP. In the physics literature, the
$N=3$ totally asymmetric case was considered by Mallick, Mallick and
Rajewsky \cite{MMR99} and a matrix product steady state was
determined. It was shown that the matrices obeyed more complicated
relations than the quadratic algebra of the two species case.

In the probabilistic literature Ferrari and Martin \cite{FM07}
provided a construction for the $N$-TASEP whereby the dynamics is
related to the dynamics of an $N$ line process i.e. $N$ coupled
TASEPs. The steady state for the $N$ line process has each of its
configurations equally likely. Therefore, to sample the configurations
of the $N$-TASEP according to their steady state probabilities one
picks from a uniform distribution an $N$ line configuration and
projects this onto the corresponding $N$-TASEP configuration. Ferrari
and Martin also showed that the $N$-line construction could be
interpreted as a system of $N$ queues in tandem with priority
customers.

In a recent work \cite{EFM07} it was shown how the construction of
Ferrari and Martin can be inverted and the steady state probabilities
written down as matrix products. A proof was given for the case $N=3$
and the matrices for general $N$ were written down in a hierarchical
fashion.

In the present work we generalize the solution to the partially
asymmetric case and provide the full proof of the matrix product state
for arbitrary $N$.  Let us summarize the key points of our solution.
The matrices $X_{J}^{(N)}$, appearing in the matrix product
expression for the steady state (\ref{matrix Ansatz}), depend on $J$,
the species present at the site to which the matrix corresponds, and
$N$ the number of species represented in the system. The matrices are
defined in eq. (\ref{X[a]}) in a hierarchical way: the matrices for an
$N$ species system are expressed in terms of those for an $N{-}1$
species system. This in turn means that the weights of the $N$ species
system may be expressed in terms of the weights of an $N{-}1$ species
system via a transfer matrix which is defined in eq. (\ref{T[a]}).
For the two species case ($N=2$) the matrices obey a quadratic algebra
(\ref{DE},\ref{DA},\ref{AE}) and this allows reduction relations which
relate the weights of a system of $L$ sites to those of a system of
$L-1$ sites. However, for $N>3$ the quadratic algebra is replaced by a
more complicated set of relations (\ref{eq XXhat1},\ref{eq
XXhat2},\ref{eq XXhat3}) involving additional `hat' matrices, again
defined hierarchically in equations (\ref{X0hat[a]},\ref{Xhat[a]}).
Both the matrices $X_{J}^{(N)}$ and the hat matrices
$\hat{X}_{J}^{(N)}$ are expressed recursively in terms of auxiliary
matrices $a_{JM}^{(N)}$ defined in equations
(\ref{a=0}--\ref{aN0}). These auxiliary matrices are themselves tensor
products of the four fundamental matrices $\openone$, $\delta$,
$\epsilon$, $A$ which appear in the $N=2$ solution.  The algebraic
properties of the auxiliary matrices, in particular the symmetry
relation (\ref{sym [aJM,aKM']}) and the commutation relations
(\ref{commut aa blue}--\ref{commut aa red}), are the key to the proof
of the quadratic relations obeyed by the matrices $X_{J}^{(N)}$ and
$\hat{X}_{J}^{(N)}$ which, in turn, furnish the proof of the matrix
product representation of the stationary state.

The transfer matrix mentioned above allows us to investigate how the
construction of Ferrari and Martin generalizes to the partially
asymmetric case. For TASEP, the transfer matrix implements the
Ferrari--Martin construction explicitly. However, in the system with
partial asymmetry, the queueing process interpretation does not hold
anymore and is replaced by a more general recurrence between systems
with $N$ and $N-1$ species.

The paper is structured as follows: in section \ref{section MPASEP},
we define the multispecies asymmetric exclusion process. In section
\ref{section matrix product}, we write the matrix product
representation of the stationary state of this model. In section
\ref{section transfer matrix} we give an interpretation of this matrix
product representation in terms of a transfer matrix. Then, after
writing the algebraic relations obeyed by the auxiliary matrices
related to the matrix product representation in section \ref{section
algebraaa}, we give a complete proof of the matrix product expression
in section \ref{section proofXXhat}. \ref{appendix traces delta
epsilon A} is devoted to the calculation of some traces of product of
matrices, \ref{TMproof} proves a classification of nonzero elements of
the transfer matrix, while \ref{appendix proofs K=0} contains the proof
of a special case of the identities required to ensure the matrix
product expression is valid.

\section{Multispecies ASEP}
\label{section MPASEP}
We consider the multispecies asymmetric simple exclusion process with
both forward and backward jumps on a one dimensional lattice with
periodic boundary conditions. This stochastic model is defined on a
configuration space such that each of the $L$ sites of the lattice can
be occupied by at most one particle (exclusion rule). Each particle
has a label which is an integer between $1$ to $N$, the `class' of the
particle.  (We  use the terms `class' and `species'
interchangeably.) The unoccupied sites (holes) will be considered as
particles of class $0$. The stochastic dynamics can be expressed in
terms of exchanges of particles at neighbouring sites. The transitions
which can occur depending on the classes of both particles are
\begin{eqnarray}
JK\to KJ&\mbox{\quad with rate $1$ \quad if $1\leq J<K\leq N$}\\
KJ\to JK&\mbox{\quad with rate $q$ \quad if $1\leq J<K\leq N$}\\
J0\to 0J&\mbox{\quad with rate $1$ \quad if $1\leq J\leq N$}\\
0J\to J0&\mbox{\quad with rate $q$ \quad if $1\leq J\leq N$}\;.
\end{eqnarray}\indent
All classes of particles jump to the right with rate $1$ and to the
left with rate $q$ if the destination site is empty. When two
particles of different class are on neighbouring sites, they can
exchange with rate $1$ if the particle with the smallest class is on
the left, and with rate $q$ if it is on the right. Thus, for a
particle of class $r$, all the particles of class larger than $r$
behave as holes. At this point, it might seem natural to consider
holes as particles of class $N+1$ rather than $0$. However, 
we do not adopt this convention as it would
make the expression of the stationary state more complicated in the
following.

We use the site variable $\tau_l =0,1,\ldots,N$.
If $\tau_l =0$ the site is empty; if $\tau_l = r >0 $
site $l$  contains a $r$th-class particle. Let us denote by
${\vec{\tau}}=(\tau_1,\dots,\tau_L)$, a configuration of the system.
The dynamics of the system can be encoded in a Markov matrix $M$.  The
time evolution of the probability $\prob_{t}(\vec{\tau})$ to be in a
configuration $\vec{\tau}$ at time $t$ is given by the master equation
\begin{equation}
\frac{d}{dt}\prob_{t}(\vec{\tau})=\sum_{\vec{\tau}'}M(\vec{\tau},\vec{\tau}')\prob_{t}(\vec{\tau}')\;.
\label{meq}
\end{equation}
The matrix $M$ is a $(N+1)^{L}$ by $(N+1)^{L}$ matrix which acts on
the configuration space. As the numbers of particles of each class are
conserved by the dynamics, we will restrict ourselves to a
configuration space with fixed number of particles and holes. We call
$P_{r}$ the number of particles of class $r$. The restricted
configuration space $\Omega(P_{0},P_{1}, ..., P_{N})$ has dimension
\begin{equation}
\left|\Omega\right|=\frac{L!}{P_{0}!P_{1}!...P_{N}!}\;,
\end{equation}
and the restricted Markov matrix which acts on it is $\left|\Omega\right|$ by $\left|\Omega\right|$.

\section{Matrix product formulation of the stationary state of the $N$-species ASEP}
\label{section matrix product}
The matrix product formulation was first used to solve the TASEP on a
lattice of length $L$ with open boundary conditions \cite{DEHP93}. It
was then extended to the 2-ASEP on the ring ${\mathbb Z}_L$
\cite{DJLS93}, which is our starting point. 
In this case the site  variable is $\tau_l =0,1,2$ which implies that site $l$ is
empty, contains a first-class particle or contains a second-class
particle, respectively.
For the 2-ASEP the matrix product formulation
consists of writing the
weight of a given configuration as a trace of a product of $L$
matrices corresponding to the classes of particles on the different
sites.  In the
matrix product formulation \cite{DJLS93} it has been proved that the
stationary measure may be written as
\begin{equation}
 \prob( {\vec{\tau}}) =
 \frac{1}{Z(P_0,P_1,P_2)} W({\vec{\tau}})\;,
\label{P2mat}
\end{equation}
where the weight of the configuration is given by
\begin{equation}
W({\vec{\tau}})= \mbox{Tr} \left[
  \prod_{l=1}^L X_{\tau_l} \right]
\label{weight}
\end{equation}
and Tr means the trace of the product of matrices $X_{\tau_l}$. The
normalization $Z(P_0,P_1,P_2)$ is chosen so that the sum of all the
probabilities is equal to 1 i.e. $Z(P_0,P_1,P_2)$ is the sum of
weights for configurations with the correct numbers $P_0,P_1,P_2$ of
holes and each species of particles. The matrices $X_{\tau_l}$ are
given by
\begin{equation}
X_0= E\;,\quad X_1=D\;,\quad X_2= A\;,
\label{EDA}
\end{equation}
that is: if the site is empty we write a matrix $E$; if the site
contains a first-class particle we write a matrix $D$; if the site
contains a second-class particle we write a matrix $A$. The matrices
$D,E,A$ obey the algebraic rules
\begin{eqnarray}
DE -q ED &=& (1-q)(D + E) \label{DE}\\
DA-q AD &=& (1-q) A\label{DA}\\
AE -q EA &=& (1-q)A \; . \label{AE}
\end{eqnarray}
The only remaining condition to satisfy is that representations of
this algebra may be found which give well-defined values for the
traces appearing in (\ref{P2mat}). This may be achieved for $q<1$ as
follows. Let $A$ be the diagonal, semi-infinite matrix
\begin{equation}
A=              \left(\begin{array}{ccccc}    
                  1&0&0&0&\dots\\
                  0&q&0&0&\ddots\\
                  0&0&q^2&0&\ddots\\
                  0&0&0&q^3&\ddots\\
                  \vdots &\ddots&\ddots&\ddots&\ddots
                        \end{array}  \right)\;.
\label{Arep}
\end{equation}
Then  $D$,$E$ may be chosen to be bidiagonal semi-infinite matrices
\begin{eqnarray}
D=              \left(\begin{array}{ccccc}    
                  1&\sqrt{c_0}&0&0&\dots\\
                  0&1&\sqrt{c_1}&0&\ddots\\
                  0&0&1&\sqrt{c_2}&\ddots\\
                  0&0&0&1&\ddots\\              
                  \vdots &\ddots&\ddots&\ddots&\ddots
                        \end{array}  \right)\;,
\label{Drep}\\[2ex]
E=              \left(\!\!\!\begin{array}{ccccc}    
                  1&0&0&0&\dots\\
                  \sqrt{c_0}&1&0&0&\ddots\\
                  0&\sqrt{c_1}&1&0&\ddots\\
                  0&0&\sqrt{c_2}&1&\ddots\\
                  \vdots &\ddots&\ddots&\ddots&\ddots
                        \end{array}  \right) \;,
\label{Erep}
\end{eqnarray}
where
\begin{equation}
c_n = 1-q^{n+1}\;.
\end{equation}

We note that, as $A$ has a geometric series for diagonal, the trace of
$A$ times a finite product of $D$ and $E$ matrices is not
divergent. See \cite{BE07} for further representations of
(\ref{DE}--\ref{DA}).\\ \indent In principle, explicit formulae for the
weights of each configuration may be obtained, either by using the
algebraic rules or taking an advantage of an explicit representation
of the matrices such as (\ref{Arep}--\ref{Erep}) (see \cite{DJLS93} or
the review \cite{BE07} for details of how these calculations are
performed).

In what follows, it will be useful to consider the matrices $\delta$, $\epsilon$ defined by
\begin{eqnarray}
\delta &=& D-\openone\\
\epsilon &=& E-\openone
\end{eqnarray}
which verify the algebraic relations
\begin{eqnarray}
\label{deltaepsilon}
&\delta\epsilon-q\epsilon\delta=(1-q) \openone\\
\label{deltaA}
&\delta A=qA\delta\\
\label{Aepsilon}
&A\epsilon=q\epsilon A\;.
\end{eqnarray}
 The first equation (\ref{deltaepsilon}) is the
$q$-deformed harmonic oscillator algebra (see \cite{EB02} for an introduction).
Writing $\delta$,$\epsilon$ out explicitly we have
\begin{eqnarray}
\delta =              \left(\begin{array}{ccccc}    
                  0&\sqrt{c_0}&0&0&\dots\\
                  0&0&\sqrt{c_1}&0&\ddots\\
                  0&0&0&\sqrt{c_2}&\ddots\\
                  0&0&0&0&\ddots\\              
                  \vdots &\ddots&\ddots&\ddots&\ddots
                        \end{array}  \right)\;,
\label{delrep}\\[2ex]
\epsilon=              \left(\!\!\!\begin{array}{ccccc}    
                  0&0&0&0&\dots\\
                  \sqrt{c_0}&0&0&0&\ddots\\
                  0&\sqrt{c_1}&0&0&\ddots\\
                  0&0&\sqrt{c_2}&0&\ddots\\
                  \vdots &\ddots&\ddots&\ddots&\ddots
                        \end{array}  \right) \;.
\label{epsrep}
\end{eqnarray}

In \cite{EFM07} the matrix product formulation of the stationary state
of the $N$-TASEP was presented, here we present the generalization to
the $N$-ASEP and give complete proofs that the solution indeed 
satisfies the stationarity condition.

We first fix our notation. We denote the configuration of the
system $\{ \tau_1,\ldots,\tau_L \}$ by $\vec{\tau}$.  We use
$X_{\tau_{l}}^{(N)}$ to denote the matrix associated to the state
$\tau_{l}$ of site $l$ in a system containing $N$ species of
particle. Thus, the subscript $\tau_{l} = 0,1,\ldots N$ indicates to
which species of particle the matrix is associated, and the
superscript $(N)$ indicates that there are $N$ species of particles in
the system. This is required because the matrix corresponding to a
species will vary according to the total number of species in the
system. The stationary probabilities become in the matrix product
formulation
\begin{equation}
\label{matrix Ansatz}
P(\vec{\tau})=\frac{1}{Z(P_0,\ldots, P_N)}\Tr\left[X_{\tau_{1}}^{(N)}X_{\tau_{2}}^{(N)}...X_{\tau_{L}}^{(N)}\right]\;.
\end{equation}
The matrices $X^{(N)}_{\tau_{l}}$ are given by
\begin{equation}
\label{X[a]}
X_{J}^{(N)}=\sum_{M=0}^{N-1}a_{JM}^{(N)}\otimes X_{M}^{(N-1)}\qquad\mbox{for $0\leq J\leq N$}
\end{equation}
with 
\begin{equation}
X_{0}^{(1)}=X_{1}^{(1)}=1\;.
\label{X0}
\end{equation}
Note that  we use a slightly different 
notation to  \cite{EFM07}
for the TASEP; with our  notation, some of the $a_{JM}^{(N)}$ will be
equal to zero. 

The matrices $a_{JM}^{(N)}$ are given by
\begin{eqnarray}
\label{a=0}
\!\!\!\!\!\!\!\!\!\!\!\!\!\!\!\!\!\!\!\!\!\!\!\!a_{JM}^{(N)}= 0\quad&&\mbox{for $0<M<J$}\\
\label{aJM}
\!\!\!\!\!\!\!\!\!\!\!\!\!\!\!\!\!\!\!\!\!\!\!\!a_{JM}^{(N)}= A^{\otimes(J-1)}\otimes\delta\otimes\openone^{\otimes(M-J-1)}\otimes\epsilon\otimes\openone^{\otimes(N-M-1)}\quad&&\mbox{for $0<J<M\leq N-1$}\\
\label{a0M}
\!\!\!\!\!\!\!\!\!\!\!\!\!\!\!\!\!\!\!\!\!\!\!\!a_{0M}^{(N)}=\openone^{\otimes(M-1)}\otimes\epsilon\otimes\openone^{\otimes(N-M-1)}\quad&&\mbox{for
$0<M\leq N-1$}\\
\label{aJJ}
\!\!\!\!\!\!\!\!\!\!\!\!\!\!\!\!\!\!\!\!\!\!\!\!a_{JJ}^{(N)}=A^{\otimes(J-1)}\otimes\openone^{\otimes(N-J)}\quad&&\mbox{for $0<J\leq N-1$}\\
\label{a00}
\!\!\!\!\!\!\!\!\!\!\!\!\!\!\!\!\!\!\!\!\!\!\!\!a_{00}^{(N)}=\openone^{\otimes(N-1)}\quad&&\mbox{for $0<J<M\leq N-1$}\\
\label{aJ0}
\!\!\!\!\!\!\!\!\!\!\!\!\!\!\!\!\!\!\!\!\!\!\!\!a_{J0}^{(N)}=A^{\otimes(J-1)}\otimes\delta\otimes\openone^{\otimes(N-J-1)}\quad&&\mbox{for $0<J\leq N-1$}\\
\label{aN0}
\!\!\!\!\!\!\!\!\!\!\!\!\!\!\!\!\!\!\!\!\!\!\!\!a_{N0}^{(N)}=A^{\otimes(N-1)}\;.&&
\end{eqnarray}
Thus the matrices $X_K^{(N)}$ at level $N$ are composed of tensor
products of $\mbox{dim}_{N} = {{N} \choose 2}$ fundamental matrices
$\delta$, $\epsilon$, $A$, or $\openone$. For $N=1$, $\mbox{dim}_{N}
=0$ which implies the use of scalars for the single species case
(\ref{X0}) (i.e. all configurations are equally likely). For $N=2$,
$\mbox{dim}_{N} =1$, which implies the use of matrices as given by
(\ref{Arep}--\ref{Erep}).  Let us now check that the 2-ASEP
matrices are recovered.  We find from equations~(\ref{a=0}),
(\ref{a0M}), (\ref{aJJ}) and (\ref{aN0}), that $a_{00}^{(2)} =
\openone$, $a_{01}^{(2)} = \epsilon$, $ a_{10}^{(2)} = \delta $, $
a_{11}^{(2)} = \openone$, and $ a_{20}^{(2)} = A$. Then, from the
recursion relation~(\ref{X[a]}) we deduce that
\begin{eqnarray}
  &X_0^{(2)} =  a_{00}^{(2)} + a_{01}^{(2)} = \openone + \epsilon = E,  
  \,\,\, 
  \,\,\,   X_1^{(2)} =  a_{10}^{(2)} +  a_{11}^{(2)} = \openone + \delta = D, 
  \,\,\,\, \label{ma201}\\
  \,\,\,& \hbox{ and }    X_2^{(2)} = a_{20}^{(2)}  =  A \, ,\label{ma22}
 \end{eqnarray}
as expected.

Let us now write out the case $N = 3$. Since
$\mbox{dim}_{3} =3$, the $X_{K}^{(3)}$ matrices are built with tensor
products of three fundamental matrices. From the definitions (\ref{a=0}--\ref{aN0}), 
we have $a_{00}^{(3)}=\openone\otimes\openone$, $a_{01}^{(3)}=\epsilon\otimes\openone$,
$a_{02}^{(3)}=\openone\otimes\epsilon$, $a_{10}^{(3)}=\delta\otimes\openone$,
$a_{11}^{(3)}=\openone\otimes\openone$, $a_{12}^{(3)}=\delta\otimes\epsilon$,
$a_{20}^{(3)}=A\otimes\delta$, $a_{22}^{(3)}=A\otimes\openone$ and
$a_{30}^{(3)}=A\otimes A$. Using the recursion relation~(\ref{X[a]}) and the
results $X_0^{(2)}=E$, $X_1^{(2)}=D$, $X_2^{(2)}=A$, we obtain
\begin{eqnarray}
X_0^{(3)} &=&  a_{00}^{(3)}\otimes X_0^{(2)}+
a_{01}^{(3)}\otimes X_1^{(2)}+a_{02}^{(3)}\otimes X_2^{(2)}\nonumber \\
&=&
 \openone\otimes\openone\otimes E +  \epsilon \otimes \openone\otimes D +
 \openone\otimes \epsilon \otimes A \\[1ex]
X_1^{(3)} &=&  a_{10}^{(3)}\otimes X_0^{(2)}+
a_{11}^{(3)}\otimes X_1^{(2)}+a_{12}^{(3)}\otimes X_2^{(2)}\nonumber \\
 &=& \delta \otimes \openone\otimes E + \openone\otimes\openone\otimes D + 
 \delta \otimes \epsilon \otimes A  \\[1ex]
X_2^{(3)} &=&  a_{20}^{(3)}\otimes X_0^{(2)}+
a_{22}^{(3)}\otimes X_2^{(2)} =  A \otimes \delta \otimes E + A \otimes \openone\otimes A \\[1ex]
X_3^{(3)} &=&  a_{30}^{(3)}\otimes X_0^{(2)}
     = A \otimes A  \otimes E\, .
  \end{eqnarray}
We notice in this example that the matrices $X_{K}^{(N)}$ have the
same expression in terms of $\delta$, $\epsilon$ and $A$ as for TASEP
\cite{EFM07}; the only difference lies in the deformation by $q$ of
the algebra (\ref{deltaepsilon}--\ref{Aepsilon}) between $\delta$,
$\epsilon$ and $A$.

In later sections we shall establish the
algebraic rules, satisfied by the matrices $X^{(N)}_{\tau_{l}}$, which
generalize the quadratic algebra (\ref{DE}--\ref{DA}) of the $N=2$
case. First we discuss how the recursive structure (\ref{X[a]}) allows
the stationary weights for an $N$-species system to be written in
terms of those for an $(N{-}1)$-species system.

\section{Transfer matrix interpretation of the matrix product representation}
\label{section transfer matrix}

In this section, we show that the matrix product representation
(\ref{matrix Ansatz}) can be conveniently rewritten in terms of a
transfer matrix acting on the configuration space.

\subsection{Definition of the Transfer Matrix}
We begin by noting that (\ref{matrix Ansatz}) and (\ref{X[a]}) define the
stationary weights for a system with $N$ species of particle
recursively in terms of the stationary weights for a system with $N-1$
species:
\begin{eqnarray}
\label{vect stat recur}
\Tr\left[X_{j_{1}}^{(N)}...X_{j_{L}}^{(N)}\right]&=&\sum_{i_{1},...,i_{L}=0}^{N-1}\Tr\left[\left(a_{j_{1}i_{1}}^{(N)}\otimes
X_{i_{1}}^{(N-1)}\right)...\left(a_{j_{L}i_{L}}^{(N)}\otimes
X_{i_{L}}^{(N-1)}\right)\right]\nonumber\\ &=&\sum_{\vec{i}}
\Tr\left[\left(a_{j_{1}i_{1}}^{(N)}\ldots
a_{\tau_{L}i_{L}}^{(N)}\right)\otimes
\left(X_{i_{1}}^{(N-1)}\ldots X_{i_{L}}^{(N-1)}\right)\right]\nonumber\\
&=&\sum_{\vec{i}}\Tr\left[a_{j_{1}i_{1}}^{(N)}...a_{j_{L}i_{L}}^{(N)}\right]\Tr\left[X_{i_{1}}^{(N-1)}...X_{i_{L}}^{(N-1)}\right]\;.
\label{reduc}
\end{eqnarray}
We use the notation $\vec{j}\equiv(j_{1},...,j_{L})$ and
$\vec{i}\equiv(i_{1},...,i_{L})$, where $\vec{j}$ is a configuration
of a system with $N$ species of particle and $\vec{i}$ is a
configuration of a system with $N-1$ species of particle.
Each $j_{l}$ can take values from $0$ to $N$ and each $i_{l}$ can take values from $0$ to $N-1$.
The sum in (\ref{reduc}) is over all configurations $\vec{i}$ with $N-1$ species.

Let us now introduce a notation for the configuration space which will be of use in the sequel.
We call $V^{(N)}$ the $N+1$ dimensional vector space corresponding to the
$N+1$ possible states of a site for a system with $N$ species. 
If we do not specify 
the number of particles of each species, the total configuration space of the system is
$V_{L}^{(N)}\equiv\left(V^{(N)}\right)^{\otimes L}$.
We denote a vector in this space corresponding to configuration
$ \vec{j}=(j_{1},...,j_{L})$  by $|\vec{j}\rangle$.
The set of all
$(N+1)^{L}$ possible configuration vectors $|\vec{j}\rangle$ form a basis of $V_{L}^{(N)}$. 
We denote  the steady state eigenvector for a system containing $N$ species of 
particle by $|N\rangle$ where
\begin{equation}
| N \rangle = \sum_{\vec{j}} W(\vec{j}) | \vec{j}\rangle\;.
\end{equation}
The stationary state weights $W(\vec{j})$ (\ref{weight}) are then
given by
\begin{equation}
W(\vec{j}) = \langle \vec{j} | N \rangle \;.
\end{equation}
Using this notation we can write relation (\ref{reduc}) in the form
\begin{equation}
\langle\vec{j}|N\rangle=\sum_{\vec{i}}\langle\vec{j}|T_{L}^{(N)}|\vec{i}\rangle\langle\vec{i}|N-1\rangle\;,
\label{Tdef}
\end{equation}
where $T_{L}^{(N)}$ is the {\em transfer matrix} for a system with $N$ species.
The matrix element $\langle  \vec{j} |T_{L}^{(N)}| \vec{i} \rangle$
can be thought of as  representing a transition from configuration
$\vec{i}$ in $V_{L}^{(N-1)}$ to configuration $\vec{j}$ in $V_{L}^{(N)}$.
(This transition is, of course, not the same as a dynamical transition
given by the Markov matrix.) The transfer matrix is used to express
the stationary weights for a system with $N$ species linearly in terms
of the weights for a system with $N-1$ species. We identify from
(\ref{reduc}) the elements of the transfer matrix $T_{L}^{(N)}$  as
\begin{equation}
\label{T[a]}
\langle\vec{j}|T_{L}^{(N)}|\vec{i}\rangle\equiv\Tr\left[a_{j_{1}i_{1}}^{(N)}\ldots
a_{j_{L}i_{L}}^{(N)}\right]\;.
\end{equation}
We can write the relation (\ref{Tdef})
more simply as
\begin{equation}
\label{recur eigenvector}
|N\rangle=T_{L}^{(N)}|N-1\rangle\;,
\end{equation}
and iterating we obtain
\begin{equation}
|N\rangle=T_{L}^{(N)}\ldots T_{L}^{(2)}|1\rangle\;,
\end{equation}
where the eigenvector of the system with only one species $|1\rangle$ is
such that each configuration has the same weight.

Let us now formalise the mathematical structure of $T_L^{(N)}$.
The transfer matrix $T_{L}^{(N)}$ is a $(N+1)^{L}\times N^{L}$ rectangular matrix
which takes a vector  in $V_{L}^{(N-1)}$ and sends it
to  in $V_{L}^{(N)}$. 
It is expressed as a trace  of a product of the local
tensors $a_{ji}^{(N)}$.
From 
(\ref{a=0}--\ref{aN0}), we observe that these $a_{ji}^{(N)}$ are
themselves tensorial products of elements of the set
$\mathcal{F}=\{\delta,\epsilon,A,\openone\}$. Thus, the building
blocks of the transfer matrix and indeed
the matrices $X_{J}^{(N)}$ are the four infinite
matrices of the set $\mathcal{F}$. We will call $\mathcal{A}$ the
infinite dimensional space on which the elements of $\mathcal{F}$
act. The matrices $a_{JM}^{(N)}$ act on the auxiliary space
$\mathcal{A}^{(N)}\equiv\mathcal{A}^{\otimes(N-1)}$. They can be seen
as the element $JM$ ($0\leq J\leq N$ and $0\leq M\leq N-1$) of a
rectangular $(N+1)\times N$ matrix $a^{(N)}$. For example,
\begin{equation}
a^{(2)}=\left(
\begin{array}{cc}
\openone & \epsilon\\
\delta & \openone\\
A & 0
\end{array}
\right)
\quad\mbox{ and }\quad
a^{(3)}=\left(
\begin{array}{ccc}
\openone\otimes\openone & \epsilon\otimes\openone & \openone\otimes\epsilon\\
\delta\otimes\openone & \openone\otimes\openone & \delta\otimes\epsilon\\
A\otimes\delta & 0 & A\otimes\openone\\
A\otimes A & 0 & 0
\end{array}
\right)\;.
\end{equation}

\subsection{Interpretation of the $T_{L}^{(N)}$ matrices}
\label{interpTM}
In this subsection, we give an interpretation of the transfer matrices
$T_{L}^{(N)}$ in terms of forbidden and allowed transitions between
configurations of particles. 
Forbidden transitions correspond to matrix elements of
$\langle  \vec{j} |T_{L}^{(N)}| \vec{i} \rangle$ that are equal to zero.
We prove in \ref{TMproof} that the only nonzero
matrix elements $\langle\vec{j}|T_{L}^{(N)}|\vec{i}\rangle$ between an
initial configuration $\vec{i}$ of particles of species between $0$
and $N-1$ and a final configuration $\vec{j}$ of particles of species
between $0$ and $N$ are characterized by the following rules:
\begin{itemize}
\item at each site, a hole in the initial configuration $\vec{i}$ can remain a hole
or become a particle of any class between $1$ and $N$ in the final configuration $\vec{j}$.
\item at each site, a particle of class $i$ between $1$ and $N-1$ can become either a
hole or a particle of class $j$ between $1$ and $i$ (the class can only decrease).
\item there is global  conservation of the number of particles of each class between $1$ and $N-1$.
\end{itemize}
These rules are proven in \ref{TMproof}.
In particular, we note that the number of holes decreases (or stays the same)
whereas the number of particles of class $N$ can only increase from
none. Thus, the only way to create a particle of class $N$ is to
remove a hole and create a class $N$ particle instead.
The allowed local transitions for $N=2$ and $N=3$ along with the 
corresponding local tensors $a_{ji}^{(N)}$ appearing in the transfer matrix element  are illustrated in figures
\ref{fig 5transitions N=2} and \ref{fig 9transitions N=3}.
\begin{figure}
\setlength{\unitlength}{1mm}
\newcommand{\particleZero}[1]{\multiput #1(0,2){5}{\thinlines\color{black}\line(0,1){2}\color{white}\line(0,1){1}\color{black}\thinlines}}
\newcommand{\particleOne}[1]{\put #1{\thinlines\color{black}\line(0,1){10}\color{black}\thinlines}}
\newcommand{\particleTwo}[1]{\multiput #1(-0.2,0){2}{\thinlines\color{red}\line(0,1){10}\color{black}\thinlines}\multiput #1(0.2,0){2}{\thinlines\color{red}\line(0,1){10}\color{black}\thinlines}\put #1{\thinlines\color{white}\line(0,1){10}\color{black}\thinlines}}
\begin{center}
\begin{picture}(120,40)
\put(9,5){$i$}\put(10,10){\thicklines\vector(0,1){20}\thinlines}\put(12,19){$a_{ji}^{(2)}$}\put(9,32){$j$}
\put(29,5){$0$}\particleZero{(30,10)}\put(32,19){$\openone$}\particleZero{(30,20)}\put(29,32){$0$}
\put(49,5){$0$}\particleZero{(50,10)}\put(52,19){$\delta$}\particleOne{(50,20)}\put(49,32){$1$}
\put(69,5){$0$}\particleZero{(70,10)}\put(72,19){$A$}\particleTwo{(70,20)}\put(69,32){$2$}
\put(89,5){$1$}\particleOne{(90,10)}\put(92,19){$\epsilon$}\particleZero{(90,20)}\put(89,32){$0$}
\put(109,5){$1$}\particleOne{(110,10)}\put(112,19){$\openone$}\particleOne{(110,20)}\put(109,32){$1$}
\end{picture}
\end{center}
\caption{Allowed local transitions in the transfer matrix $T_{L}^{(2)}$ between a particle of class $i$ (lower row, $i=0$ or $i=1$) and a particle of class $j$ (upper row, $j=0$, $j=1$ or $j=2$) at the same site. A particle of class $0$ is represented by a vertical dashed line, a particle of class $1$ by a full line, and a particle of class $2$ by a double line. The corresponding local tensor $a_{ji}^{(2)}$ appearing in the transfer matrix element is also shown.}
\label{fig 5transitions N=2}
\end{figure}
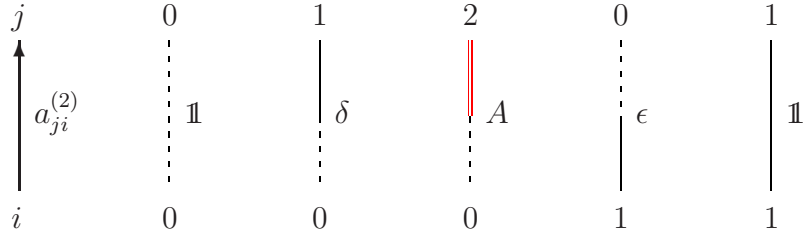
\begin{figure}
\setlength{\unitlength}{1mm}
\newcommand{\particleZero}[1]{\multiput #1(0,2){5}{\thinlines\color{black}\line(0,1){2}\color{white}\line(0,1){1}\color{black}\thinlines}}
\newcommand{\particleOne}[1]{\put #1{\thinlines\color{black}\line(0,1){10}\color{black}\thinlines}}
\newcommand{\particleTwo}[1]{\multiput #1(-0.2,0){2}{\thinlines\color{red}\line(0,1){10}\color{black}\thinlines}\multiput #1(0.2,0){2}{\thinlines\color{red}\line(0,1){10}\color{black}\thinlines}\put #1{\thinlines\color{white}\line(0,1){10}\color{black}\thinlines}}
\newcommand{\particleThree}[1]{\multiput #1(-0.4,0){2}{\thinlines\color{green}\line(0,1){10}\color{black}\thinlines}\multiput #1(0.4,0){2}{\thinlines\color{green}\line(0,1){10}\color{black}\thinlines}}
\begin{center}
\begin{picture}(100,80)
\put(9,45){$i$}\put(10,50){\thicklines\vector(0,1){20}\thinlines}\put(12,59){$a_{ji}^{(3)}$}\put(9,72){$j$}
\put(29,45){$0$}\particleZero{(30,50)}\put(32,59){$\openone\otimes\openone$}\particleZero{(30,60)}\put(29,72){$0$}
\put(49,45){$0$}\particleZero{(50,50)}\put(52,59){$\delta\otimes\openone$}\particleOne{(50,60)}\put(49,72){$1$}
\put(69,45){$0$}\particleZero{(70,50)}\put(72,59){$A\otimes\delta$}\particleTwo{(70,60)}\put(69,72){$2$}
\put(89,45){$0$}\particleZero{(90,50)}\put(92,59){$A\otimes
A$}\particleThree{(90,60)}\put(89,72){$3$}
\put(9,5){$1$}\particleOne{(10,10)}\put(12,19){$\epsilon\otimes\openone$}\particleZero{(10,20)}\put(9,32){$0$}
\put(29,5){$1$}\particleOne{(30,10)}\put(32,19){$\openone\otimes\openone$}\particleOne{(30,20)}\put(29,32){$1$}
\put(49,5){$2$}\particleTwo{(50,10)}\put(52,19){$\openone\otimes\epsilon$}\particleZero{(50,20)}\put(49,32){$0$}
\put(69,5){$2$}\particleTwo{(70,10)}\put(72,19){$\delta\otimes\epsilon$}\particleOne{(70,20)}\put(69,32){$1$}
\put(89,5){$2$}\particleTwo{(90,10)}\put(92,19){$A\otimes\openone$}\particleTwo{(90,20)}\put(89,32){$2$}
\end{picture}
\end{center}
\caption{Allowed local transitions in the transfer matrix $T_{L}^{(3)}$ between a particle of class $i$ (lower row, $i=0$, $i=1$ or $i=2$) and a particle of class $j$ (upper row, $j=0$, $j=1$, $j=2$ or $j=3$) at the same site. A particle of class $0$ is represented by a vertical dashed line, a particle of class $1$ by a  full line, a particle of class $2$ by a double line, and a particle of class $3$ by a triple line. The corresponding local tensor $a_{ji}^{(3)}$ appearing in the transfer matrix element is also shown.}
\label{fig 9transitions N=3}
\end{figure}
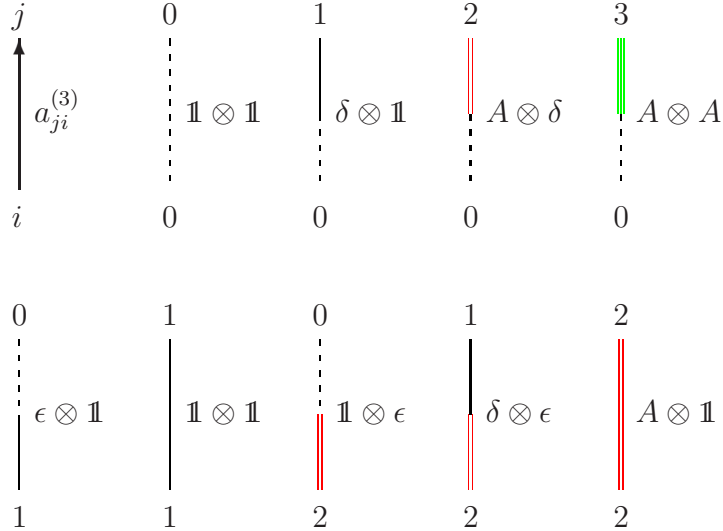

To illustrate  the utility of the transfer matrix, we work out some simple examples.
First we consider the configuration $(2,1,0)$ for $L=3$ and $N=2$.
According to the rules for nonzero elements of the transfer matrix in
the beginning of this subsection: the 2 at site 1 must have come from a 0; the
1 at site 2 could have come from a 1 or 0, the 0 at site 3 could have
come from a 1 or 0. Then the global constraint of a conserved number
of 1 implies that the only 1-species configurations which have
transitions to $(2,1,0)$ are $(0,1,0)$ and $(0,0,1)$. Using
Figure~\ref{fig 5transitions N=2} to construct the transfer matrix
elements from the local tensors 
$a_{ji}^{(N)}$, we find 
\begin{eqnarray}
W(2,1,0) &=& W(0,1,0) \Tr \left[ a_{20}^{(2)} a_{11}^{(2)} a_{00}^{(2)} \right]
+W(0,0,1) \Tr \left[ a_{20}^{(2)} a_{10}^{(2)} a_{01}^{(2)} \right]\\
&=&  \Tr \left[ A  \right]
+ \Tr \left[ A \delta \epsilon \right]\;,
\end{eqnarray}
where we have set $W(0,1,0)=W(0,0,1)=1$ (uniform measure for $N=1$).
This is to be compared with the known expression given by
the matrices (\ref{EDA})
\begin{eqnarray}
W(2,1,0) &=&  \Tr \left[ A D E  \right] =
\Tr \left[ A (\openone + \delta)(\openone + \epsilon)  \right]=
 \Tr \left[ A  + A \delta \epsilon \right]
\end{eqnarray}
where we have used the property that $\delta$ and $\epsilon$ must appear in equal numbers to have a non zero trace.

As another example we consider the configuration $(3,2,1,0)$ for $L=4$
and $N=3$.  According to the rules for nonzero elements of the
transfer matrix:  the 3 at site 1 must have come
from a 0; the 2 at site 2 could have come from a 2 or 0, the 1 at site
3 could have come from a 2,1 or 0 and the 0 at site 4 could have come
from a 2,1 or 0. Then the global constraints of conserved numbers of 1
and 2 implies that the only 2-species configurations which have
transitions to $(3,2,1,0)$ are $(0,2,1,0)$, $(0,2,0,1)$, $(0,0,2,1)$,
$(0,0,1,2)$. Using Figure~\ref{fig 9transitions N=3} to construct the
transfer matrix elements yields
\begin{eqnarray}
\lefteqn{W(3,2,1,0) =    }\\ &&
\quad W(0,2,1,0) \Tr \left[ a_{30}^{(3)} a_{22}^{(3)} a_{11}^{(3)} a_{00}^{(3)} \right]
+W(0,2,0,1) \Tr \left[ a_{30}^{(3)} a_{22}^{(3)} a_{10}^{(3)} a_{01}^{(3)} \right]\nonumber \\
&&+  W(0,0,2,1) \Tr \left[ a_{30}^{(3)} a_{20}^{(3)} a_{12}^{(3)} a_{01}^{(3)} \right]
+W(0,0,1,2) \Tr \left[ a_{30}^{(3)} a_{22}^{(3)} a_{11}^{(3)} a_{02}^{(3)} \right]\nonumber \\[1ex]
  &=& W(0,2,1,0) \Tr \left[A^2\right]\Tr \left[A\right]
+W(0,2,0,1) \Tr \left[A^2 \delta \epsilon\right]\Tr \left[A\right] \nonumber \\
&& + W(0,0,2,1) \Tr \left[A^2 \delta \epsilon\right]\Tr \left[A\delta \epsilon\right] 
  + W(0,0,1,2) \Tr \left[A^2\right]\Tr \left[A\delta \epsilon \right]\;.\nonumber
\end{eqnarray}
We used the fact that the trace of a tensorial product is equal to the product of the traces.

\subsection{Finiteness of the Trace Operation}

We have so far avoided the important question of the finiteness of the
matrix product representation (\ref{matrix Ansatz}). Using previous
results of this section and the appendices, we now prove that if there is at least one
particle of each species, the expression (\ref{matrix Ansatz}) for the
stationary measure of the $N$ species ASEP is finite. (In the case
where there are zero particles of some species, this species can be
removed from the problem by studying the corresponding system with
$N-1$ species instead of $N$.) We will focus on the case $N=3$. For
any configuration $\vec{k}$ with $3$ species, the stationary weight
is given by
\begin{equation}
W( \vec{k}) = \langle\vec{k}|3\rangle=\sum_{\vec{j}}\sum_{\vec{i}}\langle\vec{k}|T_{L}^{(3)}|\vec{j}\rangle\langle\vec{j}|T_{L}^{(2)}|\vec{i}\rangle\langle\vec{i}|1\rangle\;,
\end{equation}
where $\vec{i}$ is a configuration with $1$ species and $\vec{j}$ a
configuration with $2$ species. Let us assume that there is at least
one particle of each class in $\vec{k}$. Then, as $T_{L}^{(3)}$
conserves the number of particles of class $2$, the configurations
$\vec{j}$ that give a nonzero contribution to
$\langle\vec{k}|3\rangle$ are such that $\vec{j}$ has also at least
one particle of class $2$. From the characterization of the nonzero
elements of the transfer matrix, we must have that both $i_{l}=0$,
$j_{l}=2$, and $j_{l'}=0$, $k_{l'}=3$ at some sites $l$ and $l'$ between
$1$ and $L$ so that $\vec{i}$ and $\vec{j}$ give a nonzero
contribution to $\langle\vec{k}|3\rangle$. We see, using the
expression of the transfer matrix elements (\ref{T[a]}) 
and the form (\ref{aN0}) of $a^{(N)}_{N0}$ for $N=2$ and $N=3$, that there will be at least one $A$ in each trace of
$\langle\vec{j}|T_{L}^{(2)}|\vec{i}\rangle$ and
$\langle\vec{k}|T_{L}^{(3)}|\vec{j}\rangle$ contributing to
$\langle\vec{k}|3\rangle$. But one can calculate
explicitly traces of products of elements of $\mathcal{F}$ when there
is at least one $A$ in the product, and these traces are
finite (see \ref{appendix traces delta epsilon A}).
This proves that $\langle\vec{k}|3\rangle$ is finite. The
extension to arbitrary $N$ of this proof of the finiteness of the
matrix product representation (\ref{matrix Ansatz}) is
straightforward.

\subsection{Relation with Ferrari-Martin's construction for TASEP}
\label{sec:q0}
For the totally asymmetric case ($q=0$), the transfer matrices
$T_{L}^{(N)}$ encode Ferrari-Martin's multiline construction of the
stationary weights \cite{FM07}. We will focus on the $2$-species
stationary eigenstate constructed by $T_{L}^{(2)}$. We will see that
the set of pairs of configurations $\vec{i}$ and $\vec{j}$ for which
$\langle\vec{j}|T_{L}^{(N)}|\vec{i}\rangle\neq 0$ is smaller than in
the case $q\neq 0$. In order to know  for
which pairs of configurations $\vec{i}$ and $\vec{j}$  the matrix
element $\langle\vec{j}|T_{L}^{(N)}|\vec{i}\rangle$  is equal to zero,
we refer to representation (\ref{TMaprod}) of \ref{TMproof} which expresses
an  element of the transfer
matrix as a product of $N-1$ traces of products of $L$
fundamental matrices $\{\delta,\epsilon,A,\openone\}$. 
Therefore, we need to know which products of elements of
$\{\delta,\epsilon,A,\openone\}$ have a
trace equal to zero. For the case of the TASEP, i.e. $q=0$, the algebra
(\ref{deltaepsilon}), (\ref{deltaA}), (\ref{Aepsilon}) between
$\delta$, $\epsilon$ and $A$ reduces to
\begin{eqnarray}
\label{deltaepsilonTASEP}
&\delta\epsilon=\openone\\
\label{deltaATASEP}
&\delta A=0\\
\label{AepsilonTASEP}
&A\epsilon=0\;.
\end{eqnarray}
One can see that for any product of $l$ elements from $\{\delta,\epsilon\}$, $w=w_{1}...w_{l}$, $AwA\neq 0$ if and only if both of the following conditions are true:
\begin{itemize}
\item there are as many $\delta$ and $\epsilon$ in $w$.
\item for all $m$ between $1$ and $l$, there is at least as many $\epsilon$ as $\delta$ in $w_{m}...w_{l}$.
\end{itemize}
In that case, $AwA=AA = A$. For example,
$A\delta\delta\epsilon\epsilon\delta\epsilon A= A$ but
$A\delta\epsilon\epsilon\delta\delta\epsilon A=0$ because of an excess
of $\delta$ to the right.  Then, each of the traces of products of
elements from $\{\delta,\epsilon,A,\openone\}$ appearing in the
transfer matrix element (\ref{TMaprod}) will either be 0 or 1 and
consequently each transfer matrix element is either 0 or 1.  Then the
transfer matrix relation (\ref{Tdef}) becomes an expression for the
weight of a configuration in a system of $N$ species as a sum of
weights for `ancestor' configurations of a system of $N-1$ species.
The rules for selecting these ancestor configurations are precisely
those given by Ferrari and Martin.

We illustrate the equivalence for $q=0$ of the transfer matrix relation
(\ref{Tdef}) 
with the Ferrari and Martin algorithm in the case $N=2$,
studied by Angel \cite{Angel06}.
In this case (see Figure~\ref{fig 5transitions N=2})
an $\epsilon$ in the transfer matrix element corresponds to a particle of
class $1$ changing into a hole, a $\delta$ corresponds to a hole changing into
a particle of class $1$ and an $A$ to a hole changing into a particle of
class $2$. Therefore, for $q=0$, $\langle\vec{j}|T_{L}^{(N)}|\vec{i}\rangle\neq 0$ if and only
if the configurations $\vec{i}$ and $\vec{j}$ are such that one can go
from $\vec{i}$ to $\vec{j}$ by creating particles of class $2$ at some
of the unoccupied sites and by moving particles of class $1$ only to
the left, forbidding them to cross class $2$ particles: this is
precisely the pushing procedure of Angel for $2$ classes of
particles  which is a particular (2-line) case of 
the Ferrari-Martin $N$-line algorithm (see \cite{EFM07}).

As a simple example of the distinction between the $q=0$ and $q\neq 0$ cases let us consider
the configuration $(0,1,2)$ for $L=3$ and $N=2$.
Constructing  the transfer matrix
elements as before,  we find 
\begin{eqnarray}
W(0,1,2) &=& W(0,1,0) \Tr \left[ a_{00}^{(2)} a_{11}^{(2)} a_{20}^{(2)} \right]
+W(1,0,0) \Tr \left[ a_{01}^{(2)} a_{10}^{(2)} a_{20}^{(2)} \right]\nonumber \\
&=&  \Tr \left[ A  \right]
+ \Tr \left[  \epsilon \delta  A \right]\;,
\label{W(012)}
\end{eqnarray}
where we have set $W(0,1,0)=W(1,0,0)$.
In the case $q=0$, $\Tr \left[  \epsilon \delta  A \right] =0$
and there is only one contribution to $W(0,1,2)$. This concurs
with the pushing procedure for this example which results in just one
ancestor configuration  $(0,1,0)$. However, for $q \neq 0$, the second
term in (\ref{W(012)}) does contribute and
using the result of \ref{appendix traces delta epsilon A}
one finds 
$$
\Tr \left[\epsilon \delta  A \right] =
\frac{q}{1+q} \Tr \left[A \right]\;.
$$

A further important
observation for the 2-TASEP is that the matrix elements
$\langle\vec{j}|T_{L}^{(2)}|\vec{i}\rangle$ ``decouple on the
particles of class two'', that is, each element of the transfer matrix
$T_{L}^{(2)}$ factorizes as a product over all the pairs of two
consecutive particles of class $2$ in $\vec{j}$ of terms depending
only on the holes and first class particles in $\vec{i}$ and $\vec{j}$
between the particles of class $2$.
This is because for $q=0$ the matrix $A$ is a projector (\ref{Arep}).
However, this factorization property does
not hold anymore for the general case $q\neq 0$ as (\ref{Arep})
is no longer a projector. This fact  makes it more
difficult to find a combinatorial interpretation of the transfer
matrix such as Ferrari-Martin multiline construction since the hops of
the particles are less restricted than for TASEP.

\section{Quadratic algebra for the $a_{JM}^{(N)}$}
\label{section algebraaa}
From the algebra for $\delta$, $\epsilon$ and $A$ (\ref{deltaepsilon}--\ref{Aepsilon}) 
and the
explicit form of the $a_{JM}^{(N)}$ (\ref{a=0}--\ref{aN0}), many
quadratic relations for the $a_{JM}^{(N)}$ can be deduced. In the
proof of the matrix product representation, to be presented in section~\ref{section proofXXhat}, we will need two kinds of
these relations: symmetries of a quadratic function of the
$a_{JM}^{(N)}$ under the exchange of the second indices of the
$a_{JM}^{(N)}$ and relations allowing us to commute two
$a_{JM}^{(N)}$.

\subsection{Symmetry relations}
We have the following relation between commutators
\begin{equation}
\label{sym [aJM,aKM']}
\left[a_{JM}^{(N)},a_{KM'}^{(N)}\right]=\left[a_{JM'}^{(N)},a_{KM}^{(N)}\right]\qquad\mbox{for $J\neq 0$ and $K\neq 0$},\;
\end{equation}
which indicates that $\left[a_{JM}^{(N)},a_{KM'}^{(N)}\right]$ is symmetric under the exchange of $M$ and $M'$ for all $M$ and $M'$. For some values of $J$, $K$, $M$ and $M'$, we have the stronger properties
\begin{eqnarray}
\label{sym aJMaKM'}
&\!\!\!\!\!\!\!\!\!\!\!\!\!\!\!\!\!\!\!\!\!\!\!\!\!\!\!\!\!\!\!\!\!\!\!a_{JM}^{(N)}a_{KM'}^{(N)}=a_{JM'}^{(N)}a_{KM}^{(N)}\mbox{ for all $J$, $K$, and $M,M'\in\{0\}\cup[\max(J+1,K+1),N-1]$}\\
\label{sym aJMaJM'}
&\!\!\!\!\!\!\!\!\!\!\!\!\!\!\!\!\!\!\!\!\!\!\!\!\!\!\!\!\!\!\!\!\!\!\!a_{JM}^{(N)}a_{JM'}^{(N)}=a_{JM'}^{(N)}a_{JM}^{(N)}\mbox{ for all $J$, $M$ and $M'$},\;
\end{eqnarray}
which are also symmetries under the exchange between $M$ and $M'$.

\subsection{Commutation relations}
In the following, we will also need to exchange $a_{JM}^{(N)}$ and
$a_{KM'}^{(N)}$. If $J=K$, the symmetry relation (\ref{sym aJMaJM'})
can also be seen as a commutation relation for all value of $J$, $M$
and $M'$. For $J\neq K$, the exact form of the commutation relation
will depend on $M$ and $M'$. In the case $0<J<K$ we partition the
ensemble of couples $(M,M')$ for which $a_{JM}^{(N)}$ and
$a_{KM'}^{(N)}$ are defined and non-zero (that is
$M\in\{0\}\cup[J,N-1]$ and $M'\in\{0\}\cup[K,N-1]$) into $12$ sectors
as
\begin{equation}
\label{part MM'}
\!\!\!\!\!\!\!\!\!\!\!\!\!\!\!\!\!\!\!\!\!\!\!\!\!\!\!\!\!\!\!\!\!\!\!\!\!\!\!\begin{array}{ccccccccc}
 & \quad & M=0 & \quad & J\leq M<K & \quad & M=K & \quad & K<M\leq N-1\\
M'=0 & & 1 & & 4 & & 7 & & 10\\
M'=K & & 2 & & 5 & & 8 & & 11\\
K<M'\leq N-1 & & 3 & & 6 & & 9 & & 12
\end{array}\;.
\end{equation}
Then, we have the following commutation relations between $a_{JM}^{(N)}$ and $a_{KM'}^{(N)}$ for $0<J<K$:
\begin{eqnarray}
\label{commut aa blue}
\!\!\!\!\!\!\!\!\!\!\!\!\!\!\!\!\!\!\!\!\!\!\!\!\!\!\!\!\!\!a_{JM}^{(N)}a_{KM'}^{(N)}=qa_{KM'}^{(N)}a_{JM}^{(N)}&\qquad\mbox{in sectors $1$, $2$, $3$, $8$, $10$, $11$ and $12$}\\
\label{commut aa green}
\!\!\!\!\!\!\!\!\!\!\!\!\!\!\!\!\!\!\!\!\!\!\!\!\!\!\!\!\!\!a_{JM}^{(N)}a_{KM'}^{(N)}=a_{KM'}^{(N)}a_{JM}^{(N)}&\qquad\mbox{in sectors $4$, $5$, $6$}\\
\label{commut aa red}
\!\!\!\!\!\!\!\!\!\!\!\!\!\!\!\!\!\!\!\!\!\!\!\!\!\!\!\!\!\!a_{JM}^{(N)}a_{KM'}^{(N)}=a_{KM'}^{(N)}a_{JM}^{(N)}&-(1-q)a_{KM}^{(N)}a_{JM'}^{(N)}\qquad\mbox{in sectors $7$ and $9$}\;.
\end{eqnarray}
In the case $K=0<J$, the commutation relations take a similar form (\ref{commut2 aa blue}--\ref{commut2 aa red}).

\section{Proof of the matrix product representation}
\label{section proofXXhat}
In this section, we prove that the matrix product expression (\ref{matrix Ansatz}) gives the stationary state eigenvector of the Markov matrix.

\subsection{`Hat' matrices}
The Markov matrix of the system with $N$ classes of particles can be written in terms of the local $(N+1)^{2}$ by $(N+1)^{2}$ matrices $M_{k,k+1}^{(N)}$ which encode the rates at which the particles hop between site $k$ and $k+1$.
\begin{equation}
M=\sum_{k=1}^{L}\openone^{\otimes(k-1)}\otimes M_{k,k+1}^{(N)}\otimes\openone^{\otimes(L-k-1)}\;.
\end{equation}
For a model for which the rates do not depend on the site such as the one we are discussing, the local matrices do not depend on the site: $M_{k,k+1}^{(N)}\equiv M_{loc}^{(N)}$. For example, for $N=2$, in the basis (11,12,10,21,22,20,01,02,00)
\begin{equation}
M_{loc}^{(2)}=\left(\begin{array}{ccccccccc}
\cdot & \cdot & \cdot & \cdot & \cdot & \cdot & \cdot & \cdot & \cdot\\
\cdot & -1 & \cdot & q & \cdot & \cdot & \cdot & \cdot & \cdot\\
\cdot & \cdot & -1 & \cdot & \cdot & \cdot & q & \cdot & \cdot\\
\cdot & 1 & \cdot & -q & \cdot & \cdot & \cdot & \cdot & \cdot\\
\cdot & \cdot & \cdot & \cdot & \cdot & \cdot & \cdot & \cdot & \cdot\\
\cdot & \cdot & \cdot & \cdot & \cdot & -1 & \cdot & q & \cdot\\
\cdot & \cdot & 1 & \cdot & \cdot & \cdot & -q & \cdot & \cdot\\
\cdot & \cdot & \cdot & \cdot & \cdot & 1 & \cdot & -q & \cdot\\
\cdot & \cdot & \cdot & \cdot & \cdot & \cdot & \cdot & \cdot & \cdot
\end{array}\right)\;.
\end{equation}
The dots represent matrix elements equal to zero. The matrix product expression (\ref{matrix Ansatz}) gives the stationary state eigenvector if for all configuration $\vec{j}$ with a particle of class $j_{k}$ at site $k$ 
\begin{equation}
\sum_{\vec{i}}M_{\vec{j},\vec{i}}\Tr\left[X_{i_{1}}^{(N)}...X_{i_{L}}^{(N)}\right]=0\;,
\end{equation}
or in terms of the local jump matrix
\begin{equation}
\!\!\!\!\!\!\!\!\!\!\!\!\!\!\!\!\!\!\!\!\!\!\!\!\sum_{k=1}^{L}\sum_{i_{k},i_{k+1}=0}^{N}\left(M_{loc}^{(N)}\right)_{j_{k}j_{k+1},i_{k}i_{k+1}}\Tr\left[X_{j_{1}}^{(N)}...X_{j_{k-1}}^{(N)}X_{i_{k}}^{(N)}X_{i_{k+1}}^{(N)}X_{j_{k+2}}^{(N)}...X_{j_{L}}^{(N)}\right]=0\;.
\end{equation}
This equation will be satisfied if there exists some additional
`hat' matrices \cite{sandow,nikolaus} $\hat{X}_{0}^{(N)}$, ..., $\hat{X}_{N}^{(N)}$ such that
\begin{equation}
\sum_{i,i'=0}^{N}\left(M_{loc}^{(N)}\right)_{jj',ii'}X_{i}^{(N)}X_{i'}^{(N)}=X_{j}^{(N)}\hat{X}_{j'}^{(N)}-\hat{X}_{j}^{(N)}X_{j'}^{(N)}\;,
\label{hat}
\end{equation}
leading to a cancellation of all the terms two by two. Knowing the
form of the local matrix $M_{loc}^{(N)}$, the previous equation
(\ref{hat}) can be rewritten as:
\begin{eqnarray}
\label{eq XXhat1}
\lefteqn{\mbox{for\quad$0\leq J\leq N$}}\nonumber\\
&&\left[X_{J}^{(N)},\hat{X}_{J}^{(N)}\right]=0\;,\\[1ex]
\lefteqn{\mbox{for} \quad 0<J<K\leq\!N\quad \mbox{or}\quad  0=K<J\leq\!N}\nonumber \\[1ex]
\label{eq XXhat2}
X_{J}^{(N)}X_{K}^{(N)}-qX_{K}^{(N)}X_{J}^{(N)}&=&
\hat{X}_{J}^{(N)}X_{K}^{(N)}- X_{J}^{(N)}\hat{X}_{K}^{(N)}\\
&=& X_{K}^{(N)}\hat{X}_{J}^{(N)}-\hat{X}_{K}^{(N)}X_{J}^{(N)}\;.
\label{eq XXhat3}
\end{eqnarray}
Again, the fact that the case $0=K<J\leq N$ is singled out comes from
our choice to give the holes the index $0$ instead of $N+1$ as the
particle hierarchy would have required. Subtracting (\ref{eq XXhat2}) from (\ref{eq XXhat3})
we also have the additional
relation valid for all values of $J$ and $K$,
\begin{equation}
\label{eq XXhat4}
\left[X_{J}^{(N)},\hat{X}_{K}^{(N)}\right]+\left[X_{K}^{(N)},\hat{X}_{J}^{(N)}\right]=0\;.
\end{equation}
This equation holds even for $J=K$ because of equation (\ref{eq
XXhat1}). It tells us that
$\left[X_{J}^{(N)},\hat{X}_{K}^{(N)}\right]$ must be antisymmetric under
the exchange between $J$ and $K$. 

In the next subsection 
we will prove by induction on  $N$ that the following $\hat{X}^{(N)}$
matrices and the $X^{(N)}$ matrices defined in equation (\ref{X[a]})
verify equations (\ref{eq XXhat1}) and (\ref{eq XXhat2}):
\begin{eqnarray}
\label{X0hat[a]}
\hat{X}_{0}^{(N)}&=-(1-q)X_{0}^{(N)}+\sum_{M=0}^{N-1}a_{0M}^{(N)}\otimes\hat{X}_{M}^{(N-1)}\\
\label{Xhat[a]}
\hat{X}_{J}^{(N)}&=\sum_{M=0}^{N-1}a_{JM}^{(N)}\otimes\hat{X}_{M}^{(N-1)}\qquad\mbox{for $1\leq J\leq N$},\;
\end{eqnarray}
with $\hat{X}_{0}^{(1)}=1$ and $\hat{X}_{1}^{(1)}=q$. The
$a_{JM}^{(N)}$ are still given by equations (\ref{a=0})--(\ref{aN0}). 
In contrast with the $X_{J}^{(N)}$ matrices (\ref{X[a]}),
the expression for the $\hat{X}_{J}^{(N)}$ in terms of the
$a_{JM}^{(N)}$ is not exactly the same as for TASEP: $q$ enters both
(\ref{X0hat[a]}) and the value of $\hat{X}_{1}^{(1)}$.

We first check the simple cases $N=1$ and $N=2$. Using
$X_{0}^{(1)}=1$, $X_{1}^{(1)}=1$, $\hat{X}_{0}^{(1)}=1$ and
$\hat{X}_{1}^{(1)}=q$, we see that relations 
(\ref{eq XXhat1}),(\ref{eq XXhat2}) and (\ref{eq XXhat3})  are trivially
satisfied for $N=1$.
For  $N=2$ we find
\begin{eqnarray*}
\hat{X}_{0}^{(2)} &=&
-(1-q) X_{0}^{(2)} + a^{(2)}_{00} \otimes \hat{X}_{0}^{(1)}+ 
a^{(2)}_{01} \otimes \hat{X}_{1}^{(1)}
\\
&=&  -(1-q) E + q \openone + \epsilon = q E -(1-q) \openone\\[1ex]
\hat{X}_{1}^{(2)} &=&
a^{(2)}_{10} \otimes \hat{X}_{0}^{(1)}+ 
a^{(2)}_{11} \otimes \hat{X}_{1}^{(1)}
\\
&=&  q \delta + \openone = q D +(1-q) \openone\\[1ex]
\hat{X}_{2}^{(2)} &=&
a^{(2)}_{10} \otimes \hat{X}_{0}^{(1)} =  q A 
\end{eqnarray*}
Clearly, because $X_0^{(2)}=E$, $X_1^{(2)}=D$ and $X_0^{(2)}=A$,
the commutation relations  (\ref{eq XXhat1}) are satisfied. One
can check, for example, that
\begin{eqnarray*}
\hat{X}_{1}^{(N)}X_{0}^{(N)}- X_{1}^{(N)}\hat{X}_{0}^{(N)}
&=& (qD + (1-q) \openone)E - D(qE -(1-q)\openone)\nonumber \\ &=& (1-q)(D+E) \\[1ex]
 X_{0}^{(N)}\hat{X}_{1}^{(N)}-\hat{X}_{0}^{(N)}X_{1}^{(N)}
&=& E(qD + (1-q) \openone) - (qE -(1-q)\openone)D\\ &=& (1-q)(D+E) 
\end{eqnarray*}
so that (\ref{eq XXhat2}) and (\ref{eq XXhat3}) are satisfied for $J=1$, $K=0$
and the other cases are similarly verified.\\\indent
In the next subsection, we prove the three
quadratic relations (\ref{eq XXhat1}), (\ref{eq XXhat2}) and (\ref{eq
XXhat3}) which are sufficient for the matrix product expression
(\ref{matrix Ansatz}) to give the stationary state eigenvector of the
Markov matrix. For each relation, we must distinguish the case where
some indices are equal to zero (which will be proved in \ref{appendix
proofs K=0}), giving six relations to prove. The proofs all have a
similar pattern. They are done by induction on $N$ and rely heavily on the
symmetry and commutation relations of section \ref{section algebraaa}.

\subsection{Proof of equation (\ref{eq XXhat1})}
If $J\neq 0$ we have
\begin{eqnarray}
\!\!\!\!\!\!\!\!\!\!\!\!\!\!\!\!\!\!\!\!\!\!\!\!\lefteqn{\left[X_{J}^{(N)},\hat{X}_{J}^{(N)}\right]=} \label{hatJJ}\\
\!\!\!\!\!\!\!\!\!\!\!\!\!\!\!\!\!\!\!\!\!\!\!\!\sum_{M=0}^{N-1}\sum_{M'=0}^{N-1}\left[\left(a_{JM}^{(N)}a_{JM'}^{(N)}\right)\otimes\left(X_{M}^{(N-1)}\hat{X}_{M'}^{(N-1)}\right)-\left(a_{JM'}^{(N)}a_{JM}^{(N)}\right)\otimes\left(\hat{X}_{M'}^{(N-1)}X_{M}^{(N-1)}\right)\right]\;.
\nonumber
\end{eqnarray}
Using the quadratic relation (\ref{sym aJMaJM'}) to exchange
$a_{JM'}^{(N)}$ and $a_{JM}^{(N)}$, we get
\begin{equation}
\left[X_{J}^{(N)},\hat{X}_{J}^{(N)}\right]=\sum_{M=0}^{N-1}\sum_{M'=0}^{N-1}\left(a_{JM}^{(N)}a_{JM'}^{(N)}\right)\otimes\left[X_{M}^{(N-1)},\hat{X}_{M'}^{(N-1)}\right]\;.\label{hatJJ2}
\end{equation}
Using the fact that $a_{JM}^{(N)}a_{JM'}^{(N)}$ is symmetric in $M$ and $M'$ (\ref{sym aJMaJM'}), and that from (\ref{eq XXhat4}) $\left[X_{M}^{(N-1)},\hat{X}_{M'}^{(N-1)}\right]$ is antisymmetric in $M$ and $M'$ by induction, we find that $\left[X_{J}^{(N)},\hat{X}_{J}^{(N)}\right]=0$.\\\indent
For $J=0$, we have an extra term 
compared to (\ref{hatJJ}):
\begin{eqnarray}
\!\!\!\!\!\!\!\!\!\!\!\!\!\!\!\!\!\!\!\!\!\!\!\!\!\!\!\!\!\!\!\!\!\!\!\!\left[X_{0}^{(N)},\hat{X}_{0}^{(N)}\right]=&\left[X_{0}^{(N)},-(1-q)X_{0}^{(N)}\right]
\label{hat00}\\
&+\sum_{M=0}^{N-1}\sum_{M'=0}^{N-1}\left[\left(a_{0M}^{(N)}a_{0M'}^{(N)}\right)\otimes\left(X_{M}^{(N-1)}\hat{X}_{M'}^{(N-1)}\right)-\left(a_{0M'}^{(N)}a_{0M}^{(N)}\right)\otimes\left(\hat{X}_{M'}^{(N-1)}X_{M}^{(N-1)}\right)\right]\;.\nonumber
\end{eqnarray}
But this extra term, being the commutator of $X_0^{(N)}$ with itself,
vanishes and (\ref{hat00}) reduces to (\ref{hatJJ2}) in the case
$J=0$ giving also $\left[X_{0}^{(N)},\hat{X}_{0}^{(N)}\right]=0$.

In the rest
of the proofs, we will use the following convention to lighten the
notation: $X_{M}^{(N-1)}$ and $\hat{X}_{M}^{(N-1)}$ will be written
respectively $X_{M}$ and $\hat{X}_{M}$, while the $a_{JM}^{(N)}$ will
just be written $a_{JM}$.

\subsection{Proof of equation (\ref{eq XXhat2}) ($0<J<K$)}
\label{subsection proof XXhat2 JK}
Let us define
\begin{equation}
\mathcal{A}=X_{J}^{(N)}X_{K}^{(N)}-qX_{K}^{(N)}X_{J}^{(N)}-\hat{X}_{J}^{(N)}X_{K}^{(N)}+X_{J}^{(N)}\hat{X}_{K}^{(N)}\;.
\end{equation}
We want to show that $\mathcal{A}=0$. We have:
\begin{eqnarray}
\!\!\!\!\!\!\!\!\!\!\!\!\!\!\!\!\!\!\!\!\!\!\!\!\!\!\!\!\!\!\!\!\!\!\!\mathcal{A}=\sum_{M\in\{0\}\cup[J,N-1]}\sum_{M'\in\{0\}\cup[K,N-1]}\!\!\!\!&[(a_{JM}a_{KM'})\otimes(X_{M}X_{M'})-q(a_{KM'}a_{JM})\otimes(X_{M'}X_{M})\\
&-(a_{JM}a_{KM'})\otimes(\hat{X}_{M}X_{M'})+(a_{JM}a_{KM'})\otimes(X_{M}\hat{X}_{M'})]\;.\nonumber
\end{eqnarray}
We will cut the double sum into $4$ parts and write
$\mathcal{A}=\mathcal{A}_{1}+\mathcal{A}_{2}+\mathcal{A}_{3}+\mathcal{A}_{4}$, gathering sectors from the partition
(\ref{part MM'}). $\mathcal{A}_{1}$ will be made of the sectors
$4$, $5$ and
$6$ of (\ref{part MM'}), $\mathcal{A}_{2}$ of the sectors
$1$, $3$, $10$
and $12$, $\mathcal{A}_{3}$ of the sectors
$2$, $7$, $9$ and
$11$, and $\mathcal{A}_{4}$ of the sector
$8$:
\begin{eqnarray}
\label{sectors1 proof XXhat2}
\mathcal{A}_{1}: & \quad M\in[J,K-1]\mbox{ and }M'\in\{0\}\cup[K,N-1]\\
\label{sectors2 proof XXhat2}
\mathcal{A}_{2}: & \quad M\in\{0\}\cup[K+1,N-1]\mbox{ and }M'\in\{0\}\cup[K+1,N-1]\\
\label{sectors3 proof XXhat2}
\mathcal{A}_{3}: & \left|\begin{array}{l}M\in\{0\}\cup[K+1,N-1]\mbox{ and }M'=K\\M=K\mbox{ and }M'\in\{0\}\cup[K+1,N-1]\end{array}\right.\\
\label{sectors4 proof XXhat2}
\mathcal{A}_{4}: & \quad M=K\mbox{ and }M'=K\;.
\end{eqnarray}
We will now show that $\mathcal{A}_{1}$, $\mathcal{A}_{2}$, $\mathcal{A}_{3}$ and $\mathcal{A}_{4}$ are all equal to zero.\\\\
We begin with $\mathcal{A}_{1}$ and use the commutation relation (\ref{commut aa green}). We get
\begin{equation}
\!\!\!\!\!\!\!\!\!\!\!\!\!\!\!\!\!\!\!\!\!\!\!\!\mathcal{A}_{1}=\sum_{M=J}^{K-1}\sum_{M'\in\{0\}\cup[K,N-1]}\left(a_{JM}a_{KM'}\right)\otimes\left(X_{M}X_{M'}-qX_{M'}X_{M}-\hat{X}_{M}X_{M'}+X_{M}\hat{X}_{M'}\right)\;.
\end{equation}
By induction, $X_{M}X_{M'}-qX_{M'}X_{M}-\hat{X}_{M}X_{M'}+X_{M}\hat{X}_{M'}$ vanishes (\ref{eq XXhat2}), and thus $\mathcal{A}_{1}=0$.\\\\
For $\mathcal{A}_{2}$, using the commutation relation (\ref{commut aa blue}), we obtain
\begin{equation}
\!\!\!\!\!\!\!\!\!\!\!\!\!\!\!\!\!\!\!\!\!\!\!\!\mathcal{A}_{2}=\sum_{M,M'\in\{0\}\cup[K+1,N-1]}\left(a_{JM}a_{KM'}\right)\otimes\left(\left[X_{M},X_{M'}\right]-\hat{X}_{M}X_{M'}+X_{M}\hat{X}_{M'}\right)\;.
\end{equation}
From (\ref{sym aJMaKM'}), $a_{JM}a_{KM'}$ is symmetric under the exchange of $M$ and $M'$ while $\left[X_{M},X_{M'}\right]$ is antisymmetric, as well as $-\hat{X}_{M}X_{M'}+X_{M}\hat{X}_{M'}$ which follows from a rewriting of (\ref{eq XXhat4}), by induction. This gives $\mathcal{A}_{2}=0$.\\\\
For $\mathcal{A}_{3}$, using the commutation relations (\ref{commut aa blue}) and (\ref{commut aa red}), we have
\begin{eqnarray}
\!\!\!\!\!\!\!\!\!\!\!\!\!\!\!\!\!\!\!\!\!\!\!\!\mathcal{A}_{3}=&\sum_{M'\in\{0\}\cup[K+1,N-1]}[\left(a_{JK}a_{KM'}\right)\otimes\left(X_{K}X_{M'}-qX_{M'}X_{K}-\hat{X}_{K}X_{M'}+X_{K}\hat{X}_{M'}\right)\nonumber\\
\!\!\!\!\!\!\!\!\!\!\!\!\!\!\!\!\!\!\!\!\!\!\!\!&\qquad\qquad\qquad\qquad\qquad\qquad\qquad\qquad-q(1-q)\left(a_{KK}a_{JM'}\right)\otimes\left(X_{M'}X_{K}\right)]\nonumber\\
\!\!\!\!\!\!\!\!\!\!\!\!\!\!\!\!\!\!\!\!\!\!\!\!&+\sum_{M\in\{0\}\cup[K+1,N-1]}\left(a_{JM}a_{KK}\right)\otimes\left(X_{M}X_{K}-X_{K}X_{M}-\hat{X}_{M}X_{K}+X_{M}\hat{X}_{K}\right)\;.
\end{eqnarray}
The first sum corresponds to setting $M=K$ and the second to $M'=K$. By induction, $X_{K}X_{M'}-qX_{M'}X_{K}-\hat{X}_{K}X_{M'}+X_{K}\hat{X}_{M'}=0$ in the first sum (\ref{eq XXhat2}). With the help of the commutation relation (\ref{commut aa blue}) in sectors $2$ and $11$ for $a_{KK}a_{JM'}$, we obtain, after renaming the dummy variable $M'$ to $M$
\begin{equation}
\!\!\!\!\!\!\!\!\!\!\!\!\!\!\!\!\!\!\!\!\!\!\!\!\mathcal{A}_{3}=\sum_{M\in\{0\}\cup[K+1,N-1]}\left(a_{JM}a_{KK}\right)\otimes\left(qX_{M}X_{K}-X_{K}X_{M}-\hat{X}_{M}X_{K}+X_{M}\hat{X}_{K}\right)\;.
\end{equation}
By induction, $qX_{M}X_{K}-X_{K}X_{M}-\hat{X}_{M}X_{K}+X_{M}\hat{X}_{K}=0$ from (\ref{eq XXhat3}), as only $M=0$ and $M>K$ contribute to the sum. $\mathcal{A}_{3}$ is therefore equal to $0$.\\\\
Finally, for $\mathcal{A}_{4}$, we have
\begin{equation}
\!\!\!\!\!\!\!\!\!\!\!\!\!\!\!\!\!\!\!\!\!\!\!\!\mathcal{A}_{4}=\left(a_{JK}a_{KK}-qa_{KK}a_{JK}\right)\otimes\left(X_{K}X_{K}\right)+\left(a_{JK}a_{KK}\right)\otimes\left(-\hat{X}_{K}X_{K}+X_{K}\hat{X}_{K}\right)\;.
\end{equation}
In the first term $a_{JK}a_{KK}-qa_{KK}a_{JK}=0$ because of the commutation relation (\ref{commut aa blue}), and in the last term $-\hat{X}_{K}X_{K}+X_{K}\hat{X}_{K}=0$ by induction because of (\ref{eq XXhat1}). Thus $\mathcal{A}_{4}=0$ which proves (\ref{eq XXhat2}) in the case $0<J<K$.

\subsection{Proof of equation (\ref{eq XXhat3}) ($0<J<K$)}
\label{subsection proof XXhat3 JK}
Let us define
\begin{equation}
\mathcal{B}=X_{J}^{(N)}X_{K}^{(N)}-qX_{K}^{(N)}X_{J}^{(N)}-X_{K}^{(N)}\hat{X}_{J}^{(N)}+\hat{X}_{K}^{(N)}X_{J}^{(N)}
\end{equation}
We want to show that $\mathcal{B}=0$. We have
\begin{eqnarray}
\!\!\!\!\!\!\!\!\!\!\!\!\!\!\!\!\!\!\!\!\!\!\!\!\!\!\!\!\!\!\!\!\!\!\!\mathcal{B}=\sum_{M\in\{0\}\cup[J,N-1]}\sum_{M'\in\{0\}\cup[K,N-1]}\!\!\!\!&[(a_{JM}a_{KM'})\otimes(X_{M}X_{M'})-q(a_{KM'}a_{JM})\otimes(X_{M'}X_{M})\\
&-(a_{KM'}a_{JM})\otimes(X_{M'}\hat{X}_{M})+(a_{KM'}a_{JM})\otimes(\hat{X}_{M'}X_{M})]\;.\nonumber
\end{eqnarray}
Again, we cut the double sum into $4$ parts and write
$\mathcal{B}=\mathcal{B}_{1}+\mathcal{B}_{2}+\mathcal{B}_{3}+\mathcal{B}_{4}$, gathering sectors from the partition
(\ref{part MM'}). We use the same sectors as in the previous proof
(\ref{sectors1 proof XXhat2}--\ref{sectors4 proof XXhat2}). At this
point, we have to calculate $\mathcal{B}_{1}$, $\mathcal{B}_{2}$, $\mathcal{B}_{3}$ and
$\mathcal{B}_{4}$. Using the same arguments as in subsection \ref{subsection
proof XXhat2 JK}, we find that they are all equal to zero, which
proves (\ref{eq XXhat3}) in the case $0<J<K$.

\section{Conclusion}
A solution for the stationary state of the multispecies TASEP
was first proposed by Ferrari and Martin in \cite{FM07}. Their
solution was expressed in terms of numbers of configurations of a
multiline queueing process. A matrix product representation of this
solution was then given in \cite{EFM07} for the case of periodic
boundary conditions, making the link with several works on one
dimensional exclusion processes in the physics literature. In this
paper, we have extended this matrix product solution to the previously
unsolved problem of the multispecies partially asymmetric exclusion
process on a ring. In this case, there is no known analogue to
Ferrari-Martin queueing construction.

The mathematical structure of the matrix product solution for the
stationary state reveals several interesting features. First, the
matrices are defined in a recursive fashion (see eq. (\ref{X[a]}))
using auxiliary matrices $a_{ij}^{(N)}$, and are ultimately built as
tensor products of the fundamental matrices
$\{\delta,\epsilon,A,\openone\}$ used in the $N=2$ solution. Second,
the matrices obey quadratic relations involving additional hat
matrices and these relations generalize the quadratic algebra of the
$N=2$ case (\ref{DE}--\ref{DA}). Such relations have only been
verified before in a few cases (see \cite{BE07} section 9 for a
discussion). In our solution the key to satisfying these conditions
lies in the algebraic properties of the auxiliary matrices
$a_{ij}^{(N)}$ which we presented in section~\ref{section algebraaa}

The recursive structure of the solution allows us to construct a
transfer matrix relating the stationary weight of a configuration with
$N$ species of particles to the weights of configurations with $N-1$
species. For the case $q=0$ the transfer matrix recovers the
algorithm of Ferrari and Martin, whereas for $q \neq 0$ the structure
is more complicated.

In this paper we have not attempted to calculate physical quantities
of interest, such as correlation functions. Such calculations begin
with the computation of the normalisation $Z(P_0,\ldots, P_N)$,
defined in (\ref{matrix Ansatz}), for all
system sizes and particle numbers. Even for the two species case this
computation is not easy and to our knowledge has only been
carried out for the totally asymmetric case \cite{DJLS93}. The transfer
matrix construction of section~\ref{section transfer matrix} may provide a formalism for the
computation of the normalisation for partial asymmetry and general
$N$.

There is a well known correspondence between
the matrix product representation for the one species ASEP with open
boundaries and the two species ASEP on a ring: the matrices
corresponding to the holes and to the first class particles are the
same in both cases, but the matrix corresponding to the second class
particles on the ring becomes associated with the boundaries for the
open system. This correspondence should be investigated further in the
case of general $N$ species systems. Another interesting extension is
the case of systems with different rates for the different classes of
particles, which contains in particular the ABC model \cite{EKKM}.

The multispecies asymmetric exclusion process shares with a small number
of statistical physics models the significant property of being
integrable. In particular, this means that its Markov matrix can be
diagonalized using the (nested) Bethe Ansatz. The matrix product
representation of the stationary state and the Bethe Ansatz \cite{D87,GS92,GM05,dGE06,DL98,
DE99,AB00,GM06,Cantini,PM08} are two of
the most used techniques used to obtain exact results for the
ASEP. Thanks to its rich structure, the solution of the multispecies
ASEP might help to understand more precisely their relationship.

\subsection*{Acknowledgments}
We thank Deepak Dhar and Pablo Ferrari for many useful discussions.

\appendix
\section{Traces of products of $\delta$, $\epsilon$ and $A$}
\label{appendix traces delta epsilon A}
In this appendix, we show that the product of elements of the set
$\mathcal{F}=\{\delta,\epsilon,A,\openone\}$ with at least one $A$ has
a nonzero trace for $q\neq 0$ if and only if there is the same number
of $\delta$ and $\epsilon$ in the product.  (The case $q=0$ has
already been dealt with in section~\ref{sec:q0}.)  We use an explicit
calculation of one of these traces.  For $q\neq 0$ we can first use
the relations (\ref{deltaA}, \ref{Aepsilon}) to rearrange any product
of elements of $\mathcal{F}$ by putting all the $A$ to the left. The
result will be proportional to  $\Tr(A^{p}w)$ where $w$ is a product of elements from
the set $\{\delta,\epsilon\}$. We call $r$ the number of $\delta$ in
$w$ and $s$ the number of $\epsilon$. Using again (\ref{deltaA}) and
(\ref{Aepsilon}) to commute all the $A$ matrices to the right, we get
\begin{equation}
\Tr(A^{p}w)=q^{p(s-r)}\Tr(wA^{p})\;.
\end{equation}
Using the cyclicity of the trace, we see that if $p\geq 1$ and $r\neq
s$, $\Tr(A^{p}w)$ must be equal to $0$. 

If $p\geq 1$ but $r=s$, the
algebra (\ref{deltaepsilon}) between $\delta$ and $\epsilon$ allows one to
rewrite $w$ as a linear combination with nonnegative coefficients (for
$q$ between $0$ and $1$) of terms of the form
$\epsilon^{r'}\delta^{r'}$ with $r'\leq r$. We want to show that
$\Tr(A^{p}\epsilon^{r'}\delta^{r'})>0$ in order to prove that
$\Tr(A^{p}w)\neq 0$ if $p\geq 1$ and $r=s$. 

Let us  define
\begin{equation}
f_{r}^{(p)}\equiv\Tr(A^{p}\delta^{r}\epsilon^{r})\;.
\end{equation}
We now show how  $f_{r}^{(p)}$ can be calculated easily by
recursion on $r$. Using first the deformed commutator  (\ref{deltaepsilon}) between
$\delta$ and $\epsilon$ once, we get
\begin{equation}
f_{r+1}^{(p)}=\Tr(A^{p}\delta^{r}(q\epsilon\delta+(1-q))\epsilon^{r})=(1-q)f_{r}^{(p)}+q\Tr(A^{p}\delta^{r}\epsilon\delta\epsilon^{r})\;.
\end{equation}
Using (\ref{deltaepsilon})  $r$ more times to push the rightmost $\delta$ to the right  we obtain
\begin{eqnarray}
f_{r+1}^{(p)}&=...=&(1-q)(1+q+...+q^{k})f_{r}^{(p)}+q^{k+1}\Tr(A^{p}\delta^{r}\epsilon^{k+1}\delta\epsilon^{r-k})\nonumber\\
&=...=&(1-q^{r+1})f_{r}^{(p)}+q^{r+1}\Tr(A^{p}\delta^{r}\epsilon^{r+1}\delta)\;.
\end{eqnarray}
Using equation (\ref{deltaA}) to commute the rightmost $\delta$ through all the $A$, we get
\begin{equation}
f_{r+1}^{(p)}=(1-q^{r+1})f_{r}^{(p)}+q^{p+r+1}f_{r+1}^{(p)}\;.
\end{equation}
We have found the recurrence relation
\begin{equation}
f_{r+1}^{(p)}=\frac{1-q^{r+1}}{1-q^{p+r+1}}f_{r}^{(p)}\;,
\end{equation}
which gives
\begin{equation}
f_{r}^{(p)}=\frac{\Tr{A^{p}}}{\qC{r+p}{p}{q}}\;.
\end{equation}
Here $\qC{a}{b}{q}$ is the q-deformed binomial coefficient (see e.g. \cite{EB02}) defined as
\begin{equation}
\qC{a}{b}{q}=\frac{[a]_{q}[a-1]_{q}...[1]_{q}}{([b]_{q}[b-1]_{q}...[1]_{q})([a-b]_{q}[a-b-1]_{q}...[1]_{q})}\stackrel{q\to 1}{\longrightarrow}{{a}\choose{b}}\;,
\end{equation}
the q-deformed numbers $[a]_{q}$ being defined by
\begin{equation}
[a]_{q}=\frac{1-q^{a}}{1-q}=1+q+...+q^{a-1}\stackrel{q\to 1}{\longrightarrow}a\;.
\end{equation}
Also, using representation (\ref{Arep}) we may evaluate
\begin{equation}
\Tr  A^p = \frac{1}{1-q^p}\;.
\end{equation}

Thus, the $f_{r'}^{(p)}$ are positive for all $r'$ up to the factor
$\Tr{A^{p}}$, which means that $\Tr(A^{p}w)$ cannot be equal to zero
if $p\geq 1$ and $r=s$. We have found a necessary and sufficient
condition for a product of elements from $\mathcal{F}$ to be different
from 0.

\section{Proof of the characterization of the Transfer Matrix}
\label{TMproof}
We now prove the characterization of the nonzero matrix elements of
the transfer matrix $T_{L}^{(N)}$, given in section~\ref{interpTM}. 
From (\ref{T[a]}),
$\langle\vec{j}|T_{L}^{(N)}|\vec{i}\rangle$ is a trace of products of
$a_{j_{l}i_{l}}^{(N)}$. Because of (\ref{a=0}), if there exists a site $l$ between
$1$ and $L$ such that $0<i_{l}<j_{l}$, then
$\langle\vec{j}|T_{L}^{(N)}|\vec{i}\rangle=0$. Thus, for all $l$, we
must have either $i_{l}=0$ or $j_{l}\leq i_{l}$ for the matrix
element of $T_{L}^{(N)}$ to be nonzero. In terms of the particles in the
configurations $\vec{i}$ and $\vec{j}$, this means that a hole in
$\vec{i}$ can become a particle of any class between $0$ and $N$ by
application of the transfer matrix, and that a particle of class
$i_{l}\geq 1$ can become a particle of class $j_{l}$ only if $0\leq j_{l}\leq
i_{l}$. For example, for $N=2$ (fig. \ref{fig 5transitions N=2}), only the
five following transitions are allowed in the transfer matrix : $0\to 0$, $0\to 1$, $0\to
2$, $1\to 0$ and $1\to 1$. The transition $1\to 2$ is forbidden. For
$N=3$ (fig. \ref{fig 9transitions N=3}), $9$ transitions are allowed:
$0\to 0$, $0\to 1$, $0\to 2$, $0\to 3$, $1\to 0$, $1\to 1$, $2\to 0$,
$2\to 1$, $2\to 2$. The transitions $1\to 2$, $1\to 3$ and $2\to 3$
are forbidden.

But this local constraint on the classes of the particles in $\vec{i}$
and $\vec{j}$ at each site $l$ does not characterize completely the nonzero
matrix elements of $T_{L}^{(N)}$: we will see that there is also a non-local
constraint on $\vec{i}$ and $\vec{j}$. We observe that the expression for the
transfer matrix $T_{L}^{(N)}$ (\ref{T[a]}) can be simplified further by noting
that the $a_{ji}^{(N)}$ are themselves tensorial products of $N-1$ elements of
$\mathcal{F}=\{\delta,\epsilon,A,\openone\}$  (\ref{a=0}--\ref{aN0}).
Introducing the notation
\begin{equation}
a_{ji}^{(N)}=a_{ji}^{(N,1)}\otimes...\otimes a_{ji}^{(N,N-1)}\;,
\end{equation}
we have
\begin{eqnarray}
a_{j_{1}i_{1}}^{(N)}...a_{j_{L}i_{L}}^{(N)}&=&\left(a_{j_{1}i_{1}}^{(N,1)}\otimes...\otimes a_{j_{1}i_{1}}^{(N,N-1)}\right)...\left(a_{j_{L}i_{L}}^{(N,1)}\otimes...\otimes a_{j_{L}i_{L}}^{(N,N-1)}\right)\nonumber\\
&=&\left(a_{j_{1}i_{1}}^{(N,1)}...a_{j_{L}i_{L}}^{(N,1)}\right)\otimes...\otimes\left(a_{j_{1}i_{1}}^{(N,N-1)}...a_{j_{L}i_{L}}^{(N,N-1)}\right)\;,
\end{eqnarray}
the $a_{ji}^{(N,k)}$ being elements of $\mathcal{F}$. Thus, we have
for the transfer matrix element
\begin{eqnarray}
\label{tau[a^(Nr)]}
\langle\vec{j}|T_{L}^{(N)}|\vec{i}\rangle&=&\Tr\left(a_{j_{1}i_{1}}^{(N,1)}...a_{j_{L}i_{L}}^{(N,1)}\right)\times...\times\Tr\left(a_{j_{1}i_{1}}^{(N,N-1)}...a_{j_{L}i_{L}}^{(N,N-1)}\right)\\
&=& \prod_{r=1}^{N-1} \Tr\left[\prod_{l=1}^L a_{j_{l}i_{l}}^{(N,r)}\right]\;. \label{TMaprod}
\end{eqnarray}
Representation (\ref{TMaprod}) shows that an  element of the transfer
matrix can be written as a product of $N-1$ traces of products of $L$
fundamental matrices $\{\delta,\epsilon,A,\openone\}$. In the following,
we shall call $\Tr\left[\prod_{l=1}^L a_{j_{l}i_{l}}^{(N,r)}\right]$ the
\textit{$r$th trace} of $\langle\vec{j}|T_{L}^{(N)}|\vec{i}\rangle$.
As $T_{L}^{(N)}$ is expressed in terms of traces of elements of $\mathcal{F}$,
we have to study which elements of $\mathcal{F}$ have zero trace and which
have nonzero trace in order to determine which are the nonzero matrix elements
of $T_{L}^{(N)}$. In \ref{appendix traces delta epsilon A} we showed, using
the quadratic algebra (\ref{deltaepsilon}--\ref{Aepsilon}), that the
trace of a product of elements of $\mathcal{F}$ with at least one $A$
is nonzero (for $q\neq 0$) if and only if the number of $\delta$ is
the same as the number of $\epsilon$ in the product. In terms of the
matrix elements of the transfer matrix $T_{L}^{(N)}$, this means that
each of the $N-1$ traces of
$\langle\vec{j}|T_{L}^{(N)}|\vec{i}\rangle$ must contain the same
number of $\delta$ and $\epsilon$ as we now show. 

\begin{figure}
\setlength{\unitlength}{1mm}
\newcommand{\particleZero}[1]{\multiput #1(0,2){5}{\thinlines\color{black}\line(0,1){2}\color{white}\line(0,1){1}\color{black}\thinlines}}
\newcommand{\particleOne}[1]{\put #1{\thinlines\color{black}\line(0,1){10}\color{black}\thinlines}}
\newcommand{\particleTwo}[1]{\multiput #1(-0.2,0){2}{\thinlines\color{red}\line(0,1){10}\color{black}\thinlines}\multiput #1(0.2,0){2}{\thinlines\color{red}\line(0,1){10}\color{black}\thinlines}\put #1{\thinlines\color{white}\line(0,1){10}\color{black}\thinlines}}
\newcommand{\particleThree}[1]{\multiput #1(-0.4,0){2}{\thinlines\color{green}\line(0,1){10}\color{black}\thinlines}\multiput #1(0.4,0){2}{\thinlines\color{green}\line(0,1){10}\color{black}\thinlines}}
\definecolor{gray}{gray}{0.5}
\newcommand{\cross}[1]{\thicklines\color{gray}\put #1{\line(-1,-2){11}}\put #1{\line(-1,2){10}}\put #1{\line(1,-2){10}}\put #1{\line(1,2){11}}\color{black}\thinlines}
\begin{center}
\begin{picture}(160,45)
\put(9,5){$i$}\put(10,10){\thicklines\vector(0,1){20}\thinlines}\put(12,19){$a_{ji}^{(4,2)}$}\put(9,32){$j$}
\put(29,5){$2$}\particleTwo{(30,10)}\put(32,19){$\epsilon$}\particleZero{(30,20)}\put(29,32){$0$}
\put(49,5){$2$}\particleTwo{(50,10)}\put(52,19){$\epsilon$}\particleOne{(50,20)}\put(49,32){$1$}
\put(69,5){$2$}\particleTwo{(70,10)}\put(72,19){$0$}\particleThree{(70,20)}\put(68,32){$3|4$}
\cross{(71,20)}
\put(89,5){$0$}\particleZero{(90,10)}\put(92,19){$\delta$}\particleTwo{(90,20)}\put(89,32){$2$}
\put(109,5){$1$}\particleOne{(110,10)}\put(112,19){$0$}\particleTwo{(110,20)}\put(109,32){$2$}
\cross{(111,20)}
\put(129,5){$3$}\particleThree{(130,10)}\put(132,19){$\delta$}\particleTwo{(130,20)}\put(129,32){$2$}
\put(149,5){$2$}\particleTwo{(150,10)}\put(152,19){$\openone$}\particleTwo{(150,20)}\put(149,32){$2$}
\end{picture}
\end{center}
\caption{All the situations involving a particle of class $r=2$ for
$N=4$. A particle of class $0$ is represented by a vertical dashed
line, a particle of class $1$ by a full line, a particle of class $2$
by a double line, and a particle of class $3$ or $4$ by a triple
line. The crossed-out diagrams correspond to situations which are
forbidden by (\ref{a=0}).}
\label{fig 7situations r=2 N=4}
\end{figure}
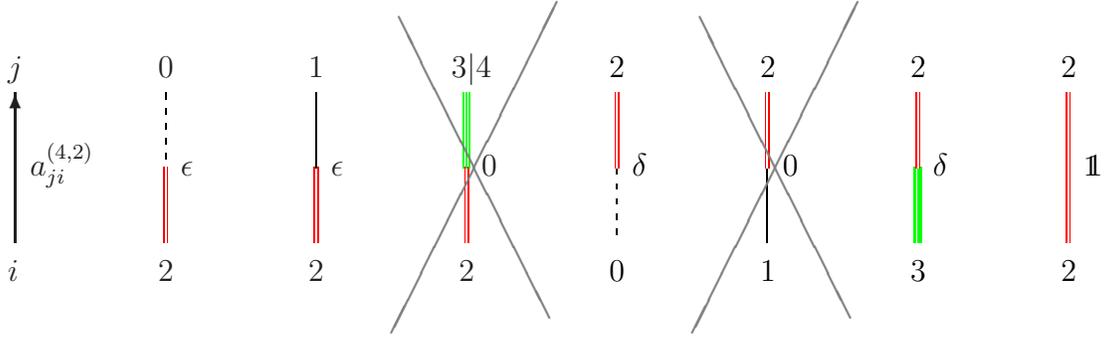

The condition of having at least
one $A$ in each trace is automatically verified if we choose the
configuration $\vec{j}$ such that it contains at least one particle of
class $N$ because a factor $a_{N0}^{(N)}$ appears (\ref{aN0}). From the expressions (\ref{a=0}--\ref{aN0}) for
the $a_{j_{l}i_{l}}^{(N)}$, we see that a $\delta$ appears in the $r$th trace
of $\langle\vec{j}|T_{L}^{(N)}|\vec{i}\rangle$ for all $l$ such that
$j_{l}=r$ and either $i_{l}=0$ or $i_{l}>r$. This corresponds to sites
at which a hole or a particle of class strictly larger than $r$ is
replaced by a particle of class $r$. On the other hand, an $\epsilon$
appears in the $r$th trace of $\langle\vec{j}|T_{L}^{(N)}|\vec{i}\rangle$
in the expression (\ref{TMaprod}) for all $l$ such that
$i_{l}=r$ and $0\leq j_{l}<r$. This corresponds to sites at which a
particle of class $r$ becomes a hole or a particle of class strictly
lower than $r$. For example for $N=2$, $\delta$ corresponds to $0\to
1$ and $\epsilon$ to $1\to 0$ (see fig. \ref{fig 5transitions
N=2}). For $N=3$, a $\delta$ appears in the first trace for $0\to 1$
and $2\to 1$ and an $\epsilon$ appears for $1\to 0$. And there is a $\delta$ in
the second trace for $0\to 2$ and an $\epsilon$ for $2\to 0$ and $2\to
1$ (see fig. \ref{fig 9transitions N=3}). In fig. \ref{fig 7situations r=2 N=4},
we draw all the transitions involving a particle of class $2$ when $N=4$
(the ones which are forbidden are crossed out), along with the corresponding
values for $a_{j_{l}i_{l}}^{(4,2)}$ (second trace). To summarize, in the $r$th trace:
\begin{itemize}
\item a $\delta$ appears when a particle of class $0$, $r+1$, $r+2$, ... , $N-1$ in $\vec{i}$ is transformed in a particle of class $r$ at the same site in $\vec{j}$.
\item an $\epsilon$ appears when a particle of class of class $r$ in $\vec{i}$ is transformed in a particle of class $0$, $1$, $2$, ... , $r-1$ at the same site in $\vec{j}$.
\end{itemize}
Therefore, the requirement that the number of $\delta$ and $\epsilon$ is
the same in the $r$th trace of $T_{L}^{(N)}$ in (\ref{TMaprod}) implies
that the number of particles of class $r$ between $1$ and $N-1$ is conserved
by the action of the transfer matrix $T_{L}^{(N)}$. But we emphasize that
neither the number of holes nor the number of particles of class $N$ is
conserved: the number of holes decreases while the number of particles of
class $N$ increases. This concludes the proof of the characterization of the
nonzero matrix elements of the transfer matrix $T_{L}^{(N)}$.

\section{Proofs of equations (\ref{eq XXhat2}) and (\ref{eq XXhat3}) for $K=0<J$}
\label{appendix proofs K=0}
The commutation relations between $a_{JM}^{(N)}$ and $a_{KM'}^{(N)}$ were given in section \ref{section algebraaa} for the case $0<J<K$. They were used in section \ref{section proofXXhat} to prove equations (\ref{eq XXhat2}) and (\ref{eq XXhat3}) for $0<J<K$. In this appendix, we will write the commutation relations between $a_{JM}^{(N)}$ and $a_{0M'}^{(N)}$ and use them to prove equations (\ref{eq XXhat2}) and (\ref{eq XXhat3}) for $K=0<J$.

\subsection{Commutation relations}
We partition the ensemble of couples $(M,M')$ for which $a_{JM}^{(N)}$ and $a_{0M'}^{(N)}$ are defined and non-zero (that is $M\in\{0\}\cup[J,N-1]$ and $M'\in[0,N-1]$) into $12$ sectors as
\begin{equation}
\label{part MM'2}
\!\!\!\!\!\!\!\!\!\!\!\!\!\!\!\!\!\!\!\!\!\!\!\!\!\!\!\!\!\!\!\!\!\!\!\!\!\!\!\begin{array}{ccccccccc}
 & \quad & M'=0 & \quad & 0<M'<J & \quad & M'=J & \quad & J<M'\leq N-1\\
M=0 & & \overline{1} & & \overline{4} & & \overline{7} & & \overline{10}\\
M=J & & \overline{2} & & \overline{5} & & \overline{8} & & \overline{11}\\
J<M\leq N-1 & & \overline{3} & & \overline{6} & & \overline{9} & & \overline{12}
\end{array}\;.
\end{equation}
Then, we have the following commutation relations between $a_{JM}^{(N)}$ and $a_{0M'}^{(N)}$:
\begin{eqnarray}
\label{commut2 aa blue}
\!\!\!\!\!\!\!\!\!\!\!\!\!\!\!\!\!\!\!\!\!\!\!\!\!\!\!\!\!\!a_{JM}^{(N)}a_{0M'}^{(N)}=qa_{0M'}^{(N)}a_{JM}^{(N)}&\qquad\mbox{in sectors $\overline{4}$, $\overline{5}$, $\overline{6}$}\\
\label{commut2 aa green}
\!\!\!\!\!\!\!\!\!\!\!\!\!\!\!\!\!\!\!\!\!\!\!\!\!\!\!\!\!\!a_{JM}^{(N)}a_{0M'}^{(N)}=a_{0M'}^{(N)}a_{JM}^{(N)}&\qquad\mbox{in sectors $\overline{1}$, $\overline{2}$, $\overline{3}$, $\overline{8}$, $\overline{10}$, $\overline{11}$ and $\overline{12}$}\\
\label{commut2 aa red}
\!\!\!\!\!\!\!\!\!\!\!\!\!\!\!\!\!\!\!\!\!\!\!\!\!\!\!\!\!\!a_{JM}^{(N)}a_{0M'}^{(N)}=qa_{0M'}^{(N)}a_{JM}^{(N)}&+(1-q)a_{0M}^{(N)}a_{JM'}^{(N)}\qquad\mbox{in sectors $\overline{7}$ and $\overline{9}$}\;.
\end{eqnarray}

\subsection{Proof of equation (\ref{eq XXhat2}) ($K=0<J$)}
\label{subsection proof XXhat2 0J}
Let us define
\begin{equation}
\overline{\mathcal{A}}=X_{J}^{(N)}X_{0}^{(N)}-qX_{0}^{(N)}X_{J}^{(N)}-\hat{X}_{J}^{(N)}X_{0}^{(N)}+X_{J}^{(N)}\hat{X}_{0}^{(N)}\;.
\end{equation}
We want to show that $\overline{\mathcal{A}}=0$. With the extra term $-(1-q)X_{0}^{(N)}$ that comes in the recursive expression for $\hat{X}_{0}^{(N)}$, the expression for $\overline{\mathcal{A}}$ rewrites
\begin{eqnarray}
\!\!\!\!\!\!\!\!\!\!\!\!\!\!\!\!\!\!\!\!\!\!\!\!\!\!\!\!\!\!\!\!\!\!\!\overline{\mathcal{A}}=\sum_{M\in\{0\}\cup[J,N-1]}\sum_{M'=0}^{N-1}&[q(a_{JM}a_{0M'})\otimes(X_{M}X_{M'})-q(a_{0M'}a_{JM})\otimes(X_{M'}X_{M})\\
&-(a_{JM}a_{0M'})\otimes(\hat{X}_{M}X_{M'})+(a_{JM}a_{0M'})\otimes(X_{M}\hat{X}_{M'})]\;.\nonumber
\end{eqnarray}
Like in the case $0<J<K$, we will cut the double sum into $4$ parts
and write
$\overline{\mathcal{A}}=\overline{\mathcal{A}}_{1}+\overline{\mathcal{A}}_{2}+\overline{\mathcal{A}}_{3}+\overline{\mathcal{A}}_{4}$,
gathering sectors from the partition (\ref{part
MM'2}). $\overline{\mathcal{A}}_{1}$ will be made of the sectors
$\overline{4}$, $\overline{5}$ and
$\overline{6}$ of (\ref{part MM'2}),
$\overline{\mathcal{A}}_{2}$ of the sectors $\overline{1}$,
$\overline{3}$, $\overline{10}$
and $\overline{12}$, $\overline{\mathcal{A}}_{3}$ of the
sectors $\overline{2}$,
$\overline{7}$, $\overline{9}$ and
$\overline{11}$, and $\overline{\mathcal{A}}_{4}$ of the
sector $\overline{8}$:
\begin{eqnarray}
\overline{\mathcal{A}}_{1}: & \quad M\in\{0\}\cup[J,N-1]\mbox{ and }M'\in[1,J-1]\\
\overline{\mathcal{A}}_{2}: & \quad M\in\{0\}\cup[J+1,N-1]\mbox{ and }M'\in\{0\}\cup[J+1,N-1]\\
\overline{\mathcal{A}}_{3}: & \left|\begin{array}{l}M=J\mbox{ and }M'\in\{0\}\cup[J+1,N-1]\\M\in\{0\}\cup[J+1,N-1]\mbox{ and }M'=J\end{array}\right.\\
\overline{\mathcal{A}}_{4}: & \quad M=J\mbox{ and }M'=J\;.
\end{eqnarray}
We will now show that $\overline{\mathcal{A}}_{1}$,
$\overline{\mathcal{A}}_{2}$, $\overline{\mathcal{A}}_{3}$ and
$\overline{\mathcal{A}}_{4}$ are all equal to zero.\\\\ We begin with
$\overline{\mathcal{A}}_{1}$ and use the commutation relation
(\ref{commut2 aa blue}). We get
\begin{equation}
\!\!\!\!\!\!\!\!\!\!\!\!\!\!\!\!\!\!\!\!\!\!\!\!\overline{\mathcal{A}}_{1}=\sum_{M=0}^{N-1}\sum_{M'=1}^{J-1}\left(a_{JM}a_{0M'}\right)\otimes\left(qX_{M}X_{M'}-X_{M'}X_{M}-\hat{X}_{M}X_{M'}+X_{M}\hat{X}_{M'}\right)\;.
\end{equation}
But, from (\ref{eq XXhat3}), $qX_{M}X_{M'}-X_{M'}X_{M}-\hat{X}_{M}X_{M'}+X_{M}\hat{X}_{M'}=0$ by induction. Thus, $\overline{\mathcal{A}}_{1}=0$.\\\\
For $\overline{\mathcal{A}}_{2}$, using the commutation relation (\ref{commut2 aa green}), we obtain
\begin{equation}
\!\!\!\!\!\!\!\!\!\!\!\!\!\!\!\!\!\!\!\!\!\!\!\!\overline{\mathcal{A}}_{2}=\sum_{M,M'\in\{0\}\cup[J+1,N-1]}\left(a_{JM}a_{0M'}\right)\otimes\left(q\left[X_{M},X_{M'}\right]-\hat{X}_{M}X_{M'}+X_{M}\hat{X}_{M'}\right)\;.
\end{equation}
From (\ref{sym aJMaKM'}), $a_{JM}a_{0M'}$ is symmetric under the exchange of $M$ and $M'$ while $\left[X_{M},X_{M'}\right]$ is antisymmetric, as well as $-\hat{X}_{M}X_{M'}+X_{M}\hat{X}_{M'}$ because of (\ref{eq XXhat4}) by induction. This gives $\overline{\mathcal{A}}_{2}=0$.\\\\
For $\overline{\mathcal{A}}_{3}$, using the commutation relations (\ref{commut2 aa green}) and (\ref{commut2 aa red}), we have
\begin{eqnarray}
\!\!\!\!\!\!\!\!\!\!\!\!\!\!\!\!\!\!\!\!\!\!\!\!\overline{\mathcal{A}}_{3}=&\sum_{M'\in\{0\}\cup[J+1,N-1]}\left(a_{JJ}a_{0M'}\right)\otimes\left(q\left[X_{J},X_{M'}\right]-\hat{X}_{J}X_{M'}+X_{J}\hat{X}_{M'}\right)\nonumber\\
\!\!\!\!\!\!\!\!\!\!\!\!\!\!\!\!\!\!\!\!\!\!\!\!&+\sum_{M\in\{0\}\cup[J+1,N-1]}[\left(a_{JM}a_{0J}\right)\otimes\left(qX_{M}X_{J}-X_{J}X_{M}-\hat{X}_{M}X_{J}+X_{M}\hat{X}_{J}\right)\nonumber\\
\!\!\!\!\!\!\!\!\!\!\!\!\!\!\!\!\!\!\!\!\!\!\!\!&\qquad\qquad\qquad\qquad\qquad\qquad\qquad\qquad+(1-q)\left(a_{0M}a_{JJ}\right)\otimes\left(X_{J}X_{M}\right)]\;.
\end{eqnarray}
By induction, $qX_{M}X_{J}-X_{J}X_{M}-\hat{X}_{M}X_{J}+X_{M}\hat{X}_{J}=0$ in the second sum (\ref{eq XXhat3}). With the help of the commutation relation (\ref{commut2 aa green}) in sectors $\overline{2}$ and $\overline{11}$ for $a_{0M}a_{JJ}$, we obtain, after renaming the dummy variable $M'$ to $M$
\begin{equation}
\!\!\!\!\!\!\!\!\!\!\!\!\!\!\!\!\!\!\!\!\!\!\!\!\overline{\mathcal{A}}_{3}=\sum_{M\in\{0\}\cup[J+1,N-1]}\left(a_{JJ}a_{0M}\right)\otimes\left(X_{J}X_{M}-qX_{M}X_{J}-\hat{X}_{J}X_{M}+X_{J}\hat{X}_{M}\right)\;.
\end{equation}
From (\ref{eq XXhat2}), $X_{J}X_{M}-qX_{M}X_{J}-\hat{X}_{J}X_{M}+X_{J}\hat{X}_{M}=0$ by induction. $\overline{\mathcal{A}}_{3}$ is then equal to $0$.\\\\
Finally, for $\overline{\mathcal{A}}_{4}$,
\begin{equation}
\!\!\!\!\!\!\!\!\!\!\!\!\!\!\!\!\!\!\!\!\!\!\!\!\overline{\mathcal{A}}_{4}=q\left(a_{JJ}a_{0J}-a_{0J}a_{JJ}\right)\otimes\left(X_{J}X_{J}\right)+\left(a_{JJ}a_{0J}\right)\otimes\left(-\hat{X}_{J}X_{J}+X_{J}\hat{X}_{J}\right)\;.
\end{equation}
In the first term $a_{JJ}a_{0J}-a_{0J}a_{JJ}=0$ because of the commutation relation (\ref{commut2 aa green}), and in the last term $-\hat{X}_{J}X_{J}+X_{J}\hat{X}_{J}=0$ by induction (\ref{eq XXhat1}). Thus $\overline{\mathcal{A}}_{4}=0$, which proves (\ref{eq XXhat2}) in the case $K=0<J$.

\subsection{Proof of equation (\ref{eq XXhat3}) ($K=0<J$)}
\label{subsection proof XXhat3 0J}
Let us define
\begin{equation}
\overline{\mathcal{B}}=X_{J}^{(N)}X_{0}^{(N)}-qX_{0}^{(N)}X_{J}^{(N)}-X_{0}^{(N)}\hat{X}_{J}^{(N)}+\hat{X}_{0}^{(N)}X_{J}^{(N)}\;.
\end{equation}
We want to show that $\overline{\mathcal{B}}=0$. With the extra term $-(1-q)X_{0}^{(N)}$ that comes in the recursive expression for $\hat{X}_{0}^{(N)}$, the expression for $\overline{\mathcal{B}}$ rewrites
\begin{eqnarray}
\!\!\!\!\!\!\!\!\!\!\!\!\!\!\!\!\!\!\!\!\!\!\!\!\!\!\!\!\!\!\!\!\!\!\!\overline{\mathcal{B}}=\sum_{M\in\{0\}\cup[J,N-1]}\sum_{M'=0}^{N-1}&[(a_{JM}a_{0M'})\otimes(X_{M}X_{M'})-(a_{0M'}a_{JM})\otimes(X_{M'}X_{M})\\
&-(a_{0M'}a_{JM})\otimes(X_{M'}\hat{X}_{M})+(a_{0M'}a_{JM})\otimes(\hat{X}_{M'}X_{M})]\;.\nonumber
\end{eqnarray}
Again, we cut the double sum into $4$ parts and write
$\overline{\mathcal{B}}=\overline{\mathcal{B}}_{1}+\overline{\mathcal{B}}_{2}+\overline{\mathcal{B}}_{3}+\overline{\mathcal{B}}_{4}$,
gathering sectors from the partition (\ref{part
MM'2}). $\overline{\mathcal{B}}_{1}$ will be made of the sectors
$\overline{4}$, $\overline{5}$ and
$\overline{6}$ of (\ref{part MM'2}),
$\overline{\mathcal{B}}_{2}$ of the sectors $\overline{1}$,
$\overline{3}$, $\overline{10}$
and $\overline{12}$, $\overline{\mathcal{B}}_{3}$ of the
sectors $\overline{2}$,
$\overline{7}$, $\overline{9}$ and
$\overline{11}$, and $\overline{\mathcal{B}}_{4}$ of the
sector $\overline{8}$. Using the same arguments as
in \ref{subsection proof XXhat2 0J} we can show that
$\overline{\mathcal{B}}_{1}$, $\overline{\mathcal{B}}_{2}$, $\overline{\mathcal{B}}_{3}$ and
$\overline{\mathcal{B}}_{4}$ are all equal to zero, which proves (\ref{eq
XXhat3}) in the case $K=0<J$.

\end{document}